%% file: nuCrayon_arXiv.tex
\documentclass[
    twocolumn,
	prd,
	amssymb,
	preprintnumbers,
	secnumarabic,
	nofootinbib]{revtex4-1}

\pdfoutput=1

\usepackage{graphicx}
\usepackage{enumitem}
\usepackage{latexsym}
\usepackage{amsfonts}
\usepackage{amssymb}
\usepackage{color}
\usepackage{amsmath}
\usepackage{slashed}
\usepackage{dcolumn}
\usepackage{verbatim}
\usepackage{float}
\usepackage{multirow}
\usepackage{xspace}
\usepackage[normalem]{ulem}
\usepackage[
pdfauthor={Nirmal Raj}]{hyperref}

\input{universalnewcommands.tex}

\def\mdm{m_{\rm DM}}
\def\GBT{G_{\rm T}}
\def\GF{G_{\rm F}}
\def\GB{G_{\rm B}}
\def\QW{Q_{\rm W}}
\def\ER{E_{\rm R}}
\def\mT{m_{\rm T}}
\def\XoneT{\acro{XENON}\osn{1}\acro{T}}

\definecolor{orange}{rgb}{1,0.5,0}
\definecolor{purple}{rgb}{1,0,1}
\definecolor{brown}{rgb}{.7,.2,.2}
\definecolor{violet}{rgb}{.6,.3,.8}
\definecolor{nicegreen}{rgb}{.3,.7,.3}

\allowdisplaybreaks

\begin{document}

\title{Monochromatic dark neutrinos and boosted dark matter in noble liquid direct detection}

\author{David McKeen}
\author{Nirmal Raj}
\affiliation{TRIUMF, 4004 Wesbrook Mall, Vancouver, BC V6T 2A3, Canada}

\begin{abstract}
If dark matter self-annihilates into neutrinos or a second component of (``boosted") dark matter that is nucleophilic, the annihilation products may be detected with high rates via coherent nuclear scattering.
A future multi-ten-tonne liquid xenon detector such as \acro{DARWIN}, and a multi-hundred-tonne liquid argon detector, \acro{argo}, would be sensitive to the flux of these particles in complementary ranges of 10--1000 MeV dark matter masses.
We derive these sensitivities after accounting for atmospheric and diffuse supernova neutrino backgrounds, and realistic nuclear recoil acceptances. 
We find that their constraints on the dark neutrino flux  may surpass neutrino detectors such as Super-Kamiokande,
and that they would extensively probe parametric regions that explain the missing satellites problem in neutrino portal models.
The \XoneT~ and \acro{Borexino} experiments currently restrict the effective baryonic coupling of thermal boosted dark matter to $\lsim 10-100 \ \times$ the weak interaction, but \acro{darwin} and \acro{argo} would probe down to couplings 10 times smaller.
Detection of boosted dark matter with baryonic couplings $\sim 10^{-3}-10^{-2} \ \times$ the weak coupling could indicate that the dark matter density profile in the centers of galactic halos become cored, rather than cuspy, through annihilations.
This work demonstrates that, alongside liquid xenon, liquid argon direct detection technology would emerge a major player in dark matter searches within and beyond the \acro{wimp} paradigm.
\end{abstract}

\maketitle

\section{Introduction}

The hunt for the identity of dark matter is a most riveting endeavor.
Particle dark matter may reveal itself 
in products of its self-annihilations, 
in target recoils in scattering experiments, or
as missing momenta in colliders.
Annihilation signals hold particular interest because they may contain information about the primordial thermal history of dark matter, and in particular, how much of its measured abundance it owes to freezing out of equilibrium in the early universe.
Conservative upper bounds on the total annihilation cross section of dark matter may be placed by constraining the flux of neutrinos, since they are the least detectable Standard Model (\acro{sm}) states \cite{Beacom:2006tt,0707.0196}.

The future of dark matter searches will depend crucially on direct detection with a panoply of noble liquid detectors, proposed with a view of achieving exposures of 
$\Oc(10)$ -- $\Oc(1000)$ tonne-years. 
These are the 
xenon-based
\acro{xenon}n\acro{t}~\cite{1512.07501},
\acro{lux}-\acro{zeplin} \cite{1509.02910} and the multi-10-ton 
\acro{darwin}~\cite{1606.07001}, 
the
argon-based
\acro{darkside}-\osn{20}\acro{K}~\cite{1707.08145},
and a multi-100-ton liquid argon detector recently christened `\acro{argo}'~\cite{largo}.
D\acro{arwin} and \acro{argo} are billed to be ``ultimate detectors" in the direct search for dark matter. 
By virtue of their large exposures, they are expected to set the best limits on the dark matter-nucleon scattering cross section, probing all the way down to the high-energy ``neutrino floor", i.e. the irreducible background of neutrino fluxes from the atmospheric scattering of cosmic rays and 
from relic supernovae.
By virtue of the large dark matter fluxes they would admit, they would also set the best limits on the dark matter mass, probing up to and beyond Planck masses \cite{1803.08044}.
In this work, we show that these experiments are also poised to become a leading probe of the neutrino flux from dark matter annihilations.
We shall henceforth call neutrinos produced in this way ``dark neutrinos".

As we will discuss below, the annihilation channel $\chi\bar\chi \ra \nu \bar\nu$ has been constrained using data from large-volume neutrino detectors \cite{0707.0196,1711.05283};
we show that \acro{darwin} and \acro{argo} sensitivities would compete with and better them in the $\sim$ 10 -- 1000 MeV dark matter mass range.
This is not entirely surprising; 
while neutrino detectors admit larger fluxes and exposures by construction, 
noble liquid direct detection experiments enjoy enhanced rates thanks to coherent scattering with large nuclei.
Moreover, search channels at neutrino detectors are typically sensitive to only $\nu_e$ and/or $\bar\nu_e$ flavors, which may make up but a fraction of the dark neutrino flux, whereas the coherent nuclear scattering channel at direct detection is equally sensitive to the flavors $\nu_e, \bar\nu_e, \nu_\mu, \bar\nu_\mu, \nu_\tau$, $\bar\nu_\tau$.
These features have been exploited to determine direct detection sensitivities to signals of neutrinos from a future core-collapse supernova burst \cite{1604.01218,1606.09243,1806.05651,1806.01417}, 
solar neutrinos \cite{1506.08309,1510.04196,Cerdeno:2016sfi,1807.07169},
and products of dark matter decay \cite{1711.04531} 
and annihilations \cite{1501.03166}.
Reference \cite{1501.03166} used a \acro{LUX} dataset with 0.027 tonne-years of exposure to constrain dark neutrinos, but we find that these constraints were weaker than those derived from neutrino experiments in \cite{0707.0196} and \cite{1711.05283};
the $\sim$ 100--1000 tonne-year datasets at \acro{darwin} and \acro{argo} would reverse this hierarchy of bounds.

Amusingly, a dark neutrino flux could produce a new neutrino floor emerging ahead of the discovery of the standard neutrino floor.
Yet, direct detection may not be able to untangle dark neutrino signals from other exotic sources of neutrinos, such as decaying dark matter \cite{1711.04531}. 
Moreover, annual modulation signals are absent.
Therefore the true discovery of dark neutrinos would entail corroborating signals at neutrino detectors, which have the ability to point to the source of the flux, in our case the galactic center.
This is an advantage when searching for electrically neutral particles such as neutrinos and photons that are produced in the annihilation of dark matter, for the propagation of charged particles would require additional astrophysical input such as the effect of magnetic fields, emission of diffuse gamma rays and synchrotron radiation, and so on.

Dark neutrino fluxes arise naturally in neutrino portal dark matter models 
\cite{
Bertoni:2014mva,
Batell:2017rol,
Batell:2017cmf}.
When interpreting our constraints in terms of this setup, parametric regions that could potentially explain the ``missing satellites" problem of structure formation can be probed extensively.
In these models the local non-relativistic population of dark matter itself scatters with nucleons in direct detection experiments, however this proceeds through a loop-induced coupling to the $Z$ boson, and the rate is suppressed.
Thus, this is an example of a theory where direct detection could find dark matter not so directly, but rather by detecting its ``friends" such as its annihilation products.
Such a detection scenario is also a generic prediction of the assisted freeze-out mechanism \cite{1112.4491}. 
In this framework, dark matter maintains thermal equilibrium with the primordial plasma even though its annihilation products do not belong to the \acro{sm}, but rather scatter efficiently with \acro{SM} states.
This possibility has given rise to a growing literature on the prospects of laboratory detection of these annihilation products, called ``boosted dark matter"
\cite{
1312.0011,
1405.7370,
1410.2246,
1411.6632,
1501.03166,
1503.02669,
1610.03486,
1611.09866,
1612.02834,
1612.06867,
1711.05278,
1712.07126,
1803.03264,
1804.07302,
1806.09154,
1811.09344}. 
While most of these efforts have focused on signals at neutrino experiments, a few such as \cite{1501.03166,1712.07126,1811.09344} have also focused on discovery at direct detection experiments.

Reference~\cite{1501.03166} in particular explored boosted dark matter with nucleophilic couplings, producing elastic nuclear recoils in direct detectors.  
We will interpret our flux sensitivities in terms of this scenario as well. 
We will first show that bounds from proton recoils in the liquid scintillator neutrino detector \acro{borexino} already outperformed those of Ref.~\cite{1501.03166} from \acro{lux}, for dark matter masses of 100--1000 MeV. 
Then we show that for present-day annihilations of dark matter with a thermal cross section, couplings as weak as the weak interaction may be probed by \acro{darwin} and \acro{argo}, improving on Ref.~\cite{1501.03166}'s limit on the couplings by a factor of $\sim$ 500.
We also argue that detecting boosted dark matter with tiny baryonic couplings may hint at a solution to the ``core-cusp" problem, and show the relevant parameter space that may be probed.

This paper is laid out as follows.
In Section~\ref{sec:reaches}, we derive the fluxes and event rates, including a careful treatment of backgrounds from atmospheric and diffuse supernova neutrinos, and nuclear recoil acceptances essential for rejecting electron recoil backgrounds from solar neutrinos.
We then translate these to projected sensitivities, and compare with bounds at neutrino detectors such as Super-Kamiokande. 
In Section~\ref{sec:interpret}, we interpret these results in terms of the neutrino portal model and boosted dark matter.
We conclude and discuss future possibilities in Section~\ref{sec:concs} .

\begin{figure*}
\includegraphics[width=.48\textwidth]{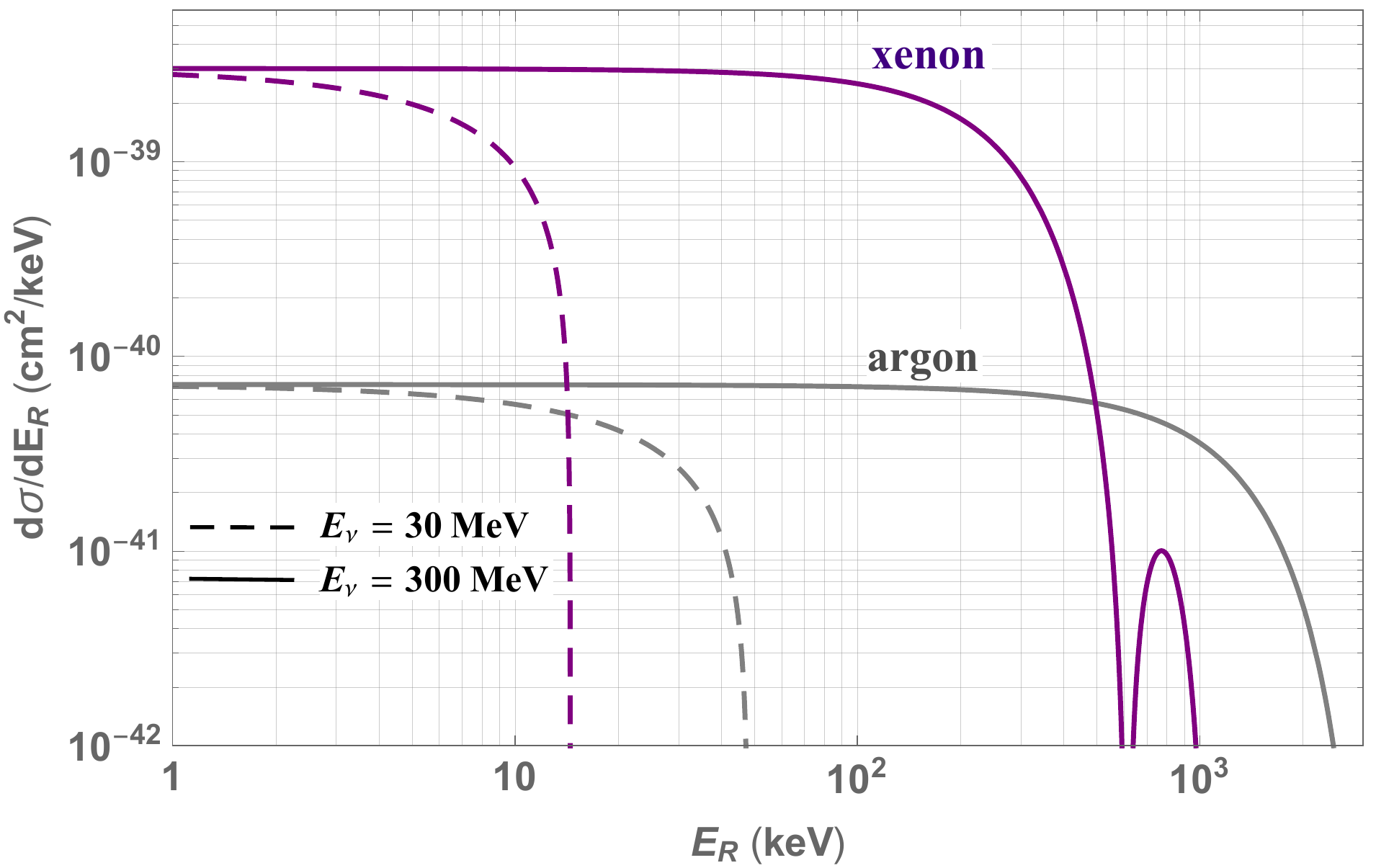} \quad
\includegraphics[width=.48\textwidth]{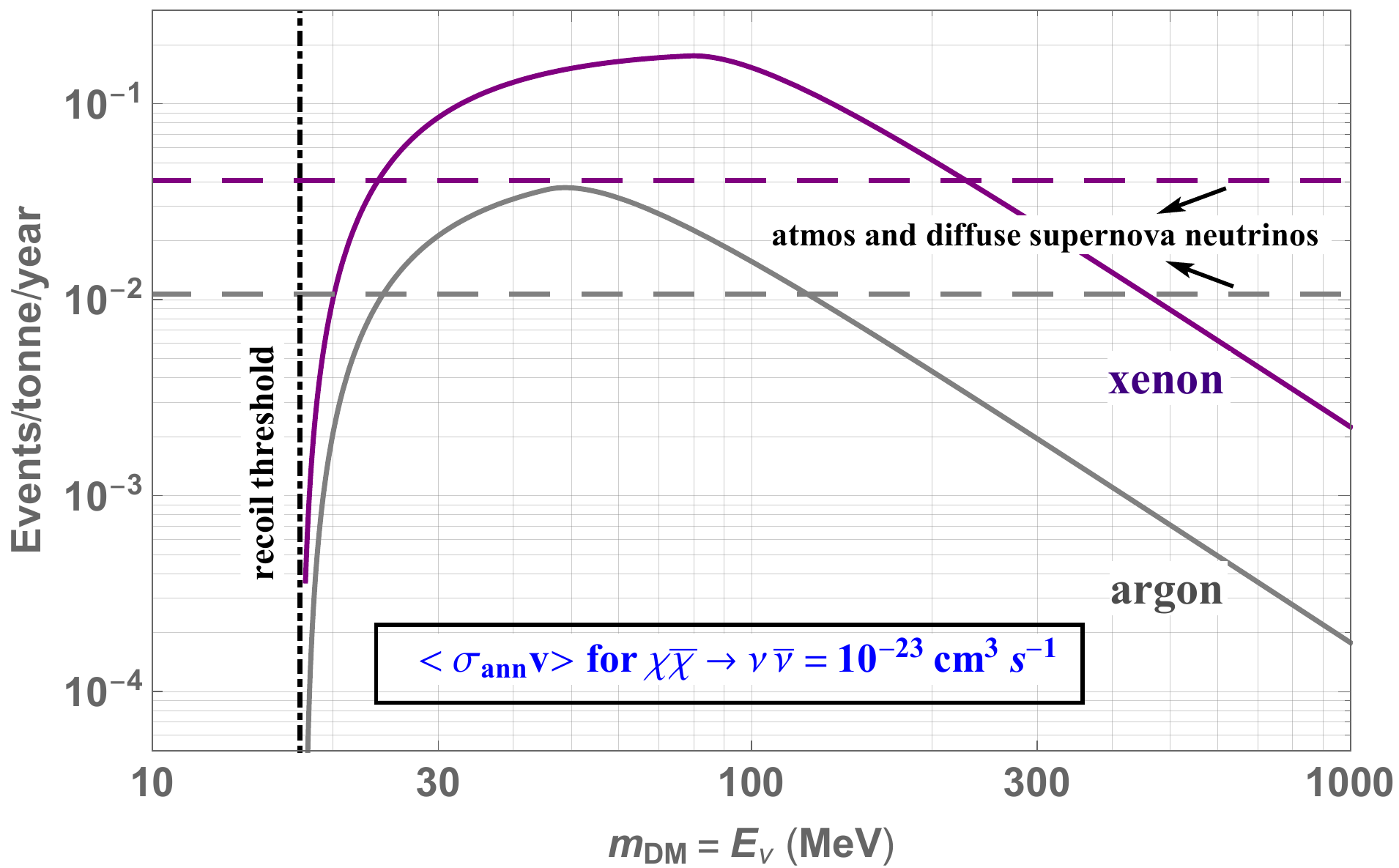}
\caption{
\textbf{{\em Left.}} Differential cross section for coherent scattering of neutrinos with xenon and argon nuclei, for neutrino energies 30 MeV and 300 MeV. 
Argon, being lighter, recoils to higher energies.
The effect of the Helm form factor, seen as a bump in the cross section, is only important at recoil energies $\sim 10^3$ keV, which far exceeds the cutoff $\sim 10^2$ keV  below which we accept events for our study.
\textbf{{\em Right.}}
The scattering rate of dark neutrinos ($\GBT$ = $\GF$ in Eq.~\eqref{eq:xsscat}) per tonne-year of exposure at liquid xenon and argon detectors, assuming dark matter annihilation cross section $\sigmaveeann = 10^{-23}$ cm$^3$/s 
and an \acro{NFW} halo profile.
Events are integrated in the recoil energy window 
5 keV -- 100 keV for xenon and 
17 keV -- 110 keV for argon;
the lower ends of these ranges correspond to neutrino energy $\simeq$ 17 MeV.
Also shown for reference is the rate of atmospheric and diffuse supernova neutrinos, which make up our primary irreducible background.
} 
\label{fig:rates}
\end{figure*}

\section{Fluxes, Signal Rates, and Sensitivities}
\label{sec:reaches}

In this section we estimate the flux of dark neutrinos and boosted dark matter from galactic and extragalactic annihilations of dark matter, which we will use to obtain scattering rates at xenon and argon detectors.
We then derive the sensitivities to these fluxes at \acro{darwin} and \acro{argo}, accounting for realistic event acceptances and irreducible neutrino backgrounds. 
We compare these with bounds from scattering on nucleons at \XoneT~and \acro{borexino}, and from other processes at neutrino experiments.

\subsection{Fluxes}
\label{subsec:flux}

The differential flux of dark neutrinos or boosted dark matter from galactic annihilations is given by
\bea
\frac{d^2\Phi}{d\Delta\Omega dE_\nu} = \eta \ \frac{r_\odot}{4\pi} \left(\frac{\rho_{{\rm DM},\odot}}{\mdm}\right)^2 \mathcal{J_{\rm ann}} \frac{\sigmaveeann}{2} \frac{dN}{dE_\nu}~,
\label{eq:flux}
\eea
where $\eta = 1/2$ accounts for dark matter not being self-conjugate,  
$r_\odot$ = 8.33 kpc is the distance of the sun from the galactic center,
the $(4\pi)^{-1}$ accounts for isotropic emission,
$\rho_{{\rm DM},\odot}$ = 0.4 GeV/cm$^3$ is the dark matter density at the solar position \cite{0907.0018},
$\mdm$ is the dark matter mass,
$\sigmaveeann$ is the thermally averaged annihilation cross section of dark matter,
and $dN/dE_\nu = 2\delta(E_\nu-\mdm)$ is the (monochromatic) energy spectrum of dark neutrinos or boosted dark matter.
The dimensionless factor $\mathcal{J_{\rm ann}}$ accounts for integrating over the dark matter distribution in the line of sight for an angular direction:
\bea
\nn \mathcal{J_{\rm ann}} = \frac{1}{r_\odot\rho^2_{{\rm DM},\odot}}\int_{\rm l.o.s.}\rho^2_{\rm DM}(\sqrt{r^2_\odot-2 s r_\odot\cos\theta +s^2})ds
\eea

We use the interpolation functions for $\mathcal{J_{\rm ann}}$ provided in \cite{1012.4515} to integrate over the $4\pi$ sky, and obtain for the Navarro-Frenk-White halo profile \cite{Navarro:1995iw}
\bea
\nn \Phi = 5.6 \times 10^{-2} \ {\rm cm}^{-2}{\rm s}^{-1} \left(\frac{\sigmaveeann}{10^{-25} \ {\rm cm}^3/{\rm s}} \right) \left(\frac{100 \ {\rm MeV}}{\mdm}\right)^2, \\
\label{eq:fluxexample}
\eea
where we have normalized $\sigmaveeann$ to the thermal cross section in this range of dark matter masses \cite{1204.3622}.

Anticipating an interpretation of our results in terms of a model where dark matter couples dominantly to the tau neutrino, we will display our main results assuming that dark neutrinos are produced exclusively through the channel $\chi\bar{\chi} \ra \nu_\tau \bar{\nu}_\tau$.
We will also briefly discuss the possibility that dark matter annihilates to neutrino mass eigenstates, $\chi\bar{\chi} \ra \nu_i \bar{\nu}_i$ ($i$=1,2,3)  with branching fractions proportional to the squared neutrino masses.
This could happen, e.g., in models where neutrino masses are generated by the breaking of lepton number symmetry. 
In either scenario, the flavor content of the dark neutrinos after propagation through astrophysical distances and arrival at earth is irrelevant for coherent scattering at direct detection experiments. 
It would, however, be important for neutrino experiments, where searches depend on the flavor of the neutrinos and/or anti-neutrinos being detected.

The total flux of dark neutrinos or boosted dark matter could also receive a contribution from dark matter annihilations in extragalactic sources. 
These may be divided into unclustered and clustured populations.
The unclustered contribution is trivially negligible: the (cosmic density)$^2$ of dark matter is $\sim 10^{-10} \ \times$ the (local density)$^2$.  
The exact contribution of the clustered population depends sensitively on the astrophysical modeling of enhancement factors from halo substructure.
For instance, using the redshift-dependent enhancement factors in \cite{1602.07282}, we find that the extragalactic flux is $\Oc(10)$ smaller than the Milky Way flux for the case of non-interacting dark matter.
For the case of dark matter interacting strongly with dark neutrinos, a scenario that we consider in Sec.~\ref{subsec:nuportal}, Ref.~\cite{1602.07282} finds that this flux is smaller by one more order of magnitude. 

However, Ref.~\cite{1809.00671} has determined that the extragalactic flux is comparable to the galactic one, by using different models of substructure enhancement.
In what follows, we will show the possible sensitivities of our detectors obtainable from such a flux.
To do so, we simply rescale our galactic flux in Eq.~\eqref{eq:fluxexample} with the total (galactic + extragalactic) flux presented in  \cite{1809.00671}.
This approach neglects redshifts in momenta, i.e. the annihilation products are still taken as monochromatic.
For this reason we do not include the extragalactic flux when displaying our main results, and only use it to visualize our optimistic sensitivities. 

Contributions to the flux may also stem from dark matter density enhancements around intermediate mass black holes \cite{Bertone:2005xz,Bertone:2006nq,Horiuchi:2006de,Brun:2007tn,Arina:2015zoa}, but we will not pursue this possibility.

\subsection{Coherent scattering rates}
\label{subsec:scatter}

We now determine the cross sections and scattering event rates of dark annihilation products at our detectors, treating the cases of dark neutrinos and boosted dark matter simultaneously.
In addition to coherent nuclear scattering, dark neutrinos also undergo electron scattering at direct detection, however the rates are many orders of magnitude smaller (see, e.g., \cite{1604.01218}), and therefore we will not consider electron recoil signals in our study.
  
For a target nucleus with mass $\mT$, 
$N$ neutrons and 
$Z$ protons,
the differential coherent scattering cross section is given by \cite{Freedman:1977xn}
\bea
\nn \frac{d\sigma}{d \ER}(E_\nu, \ER) =
\frac{\GBT^2}{4\pi} 
\QW^2 
\mT 
\left(1 - \frac{\mT \ER}{2E^2_\nu} \right)
F^2(\ER)~, \\
\label{eq:xsscat}
\eea
where 
$\QW = N - (1-4\sin^2\theta_{\rm W})Z$,
with $\sin^2\theta_{\rm W} = 0.2387$ the weak mixing angle at low energies \cite{Erler:2004in}, 
and $\GBT$ is a coupling strength that depends on both the annihilation product and target element, given by \cite{1103.3261}
\begin{align}
\displaystyle
\GBT =
 \begin{cases}
	\GF
	&\text{for dark neutrinos}~, 
	\\
	\sqrt{2}\GB ((N+Z)/\QW)
	&\text{for boosted dark matter}~.
 \end{cases}
 \label{eq:GBT}
\end{align}
Here $\GF = 1.1664 \times 10^{-5}$ GeV$^{-2}$ is the Fermi constant, and 
$\GB$ is an effective baryonic coupling of boosted dark matter, whose origins we spell out in Sec.~\ref{subsec:boosted}.  
The above equation implies that, for boosted dark matter, $\GBT/\GB = \{2.471, 2.670, 31.288\}$ for the target nuclei \{$_{~54}^{132}$Xe, $_{18}^{40}$Ar, $_{1}^{1}$H\}.

The nuclear form factor $F(\ER)$ is best parametrized by the Helm form factor \cite{Helm:1956zz} for the momentum transfers we are concerned with, and is given by 
\beq
F(\ER) =  3\frac{j_1(q r_n)}{q r_n} e^{-q^2s^2/2}~,
\eeq
where $j_1$ is the spherical Bessel function,
$s = 0.9$ fm is the nuclear skin thickness,
$q = \sqrt{2\mT\ER}$ is the momentum transfer,
and $r_n = (c^2+\frac{7}{3}\pi^2a^2-5s^2)^{1/2}$ parametrizes the nuclear radius, with $c$ and $a$ = 1.23 $A^{1/3} - 0.6$ fm and 0.52 fm respectively.

In the left panel of Fig.~\ref{fig:rates} we show $d\sigma/d\ER$ versus $\ER$ with xenon and argon targets, for neutrino energies 30 MeV and 300 MeV.
Note that the maximum recoil energy limited by kinematics is 
\beq
E_{\rm R, max}^{\rm kinem} = \frac{2E^2_\nu}{\mT + 2E_\nu}~.
\label{eq:ermaxkinem}
\eeq
 
The differential scattering rate (per tonne of detector mass) is now obtained from 
Eqs.~\eqref{eq:flux} and \eqref{eq:xsscat} as 
\beq
\frac{dR}{d\ER} =
N^{\rm ton}_{\rm T} 
\int_{E_{\nu,{\rm min}}}^\infty 
dE_\nu \frac{d\Phi}{dE_\nu} \frac{d\sigma}{d \ER}~,
\label{eq:ratescat}
\eeq
where $N^{\rm ton}_{\rm T}$ = 4.57$\times$10$^{28}$ (1.51$\times$10$^{27}$) is the number of nuclei per tonne of liquid Xe (Ar), and
$E_{\nu,{\rm min}} = \sqrt{\mT\ER/2}$ is the minimum $E_\nu$ required to induce a nuclear recoil of energy $\ER$.

In order to apply the above treatment to future noble liquid detectors, we assume the following fiducial detector masses:
\bea
\nn \acro{darwin}&:&  40 \ {\rm tonnes}, \\
\nn \acro{argo}&:& 300 \ {\rm tonnes}.
\eea

The \acro{darwin} mass is as advertised in \cite{1606.07001}, and the \acro{argo} mass is a realistic possibility \cite{private}.

To obtain the total event count at \acro{darwin}, we integrate the rate in Eq.~\eqref{eq:ratescat} over the range $\ER \in$ [5 keV, 100 keV]. 
Below this range solar neutrinos would populate a steep ``wall" of background events \cite{0903.3630,1307.5458}; 
the upper end of this range is chosen for showing conservative limits.
At \acro{argo}, we use the range $\ER \in$ [17 keV, 110 keV], where the lower end is once again chosen to evade the solar $\nu$ background, and the upper end is chosen to approximately match with the \acro{darwin} range.
The choice of these $\ER$ ranges also ensures that the contribution of inelastic processes (quasielastic scattering, production of resonant states and deep inelastic scattering) to the total event rate is negligible \cite{1307.5458}. 

In the right panel of Fig.~\ref{fig:rates} we show as a function of $\mdm$ the integrated event rate per tonne-year for scattering of dark neutrinos 
(i.e., setting $\GBT = \GF$) 
with a flux corresponding to $\sigmaveeann = 10^{-23}$ cm$^3$/s.
Also shown as horizontal dashed lines are the integrated rates of atmospheric and diffuse supernova neutrino scattering taken from \cite{0903.3630}, which constitute our main background.
The signal rate at argon detectors peaks at lower neutrino energies compared to xenon detectors since $E_{\rm R, max}^{\rm kinem}$ in Eq.~\eqref{eq:ermaxkinem} attains the maximum $\ER$ imposed by us here at a lower $E_\nu$ for argon than for xenon.
We will find below that this allows \acro{darwin} and \acro{argo} to probe some regions in complementary ranges of dark matter mass.
Figure~\ref{fig:rates} also shows that the signal rate at xenon detectors is roughly an order of magnitude higher than at argon detectors, implying that the latter require $\sim$ 10 $\times$ the exposure of the former to achieve comparable sensitivities.

\begin{figure*}
\includegraphics[width=.48\textwidth]{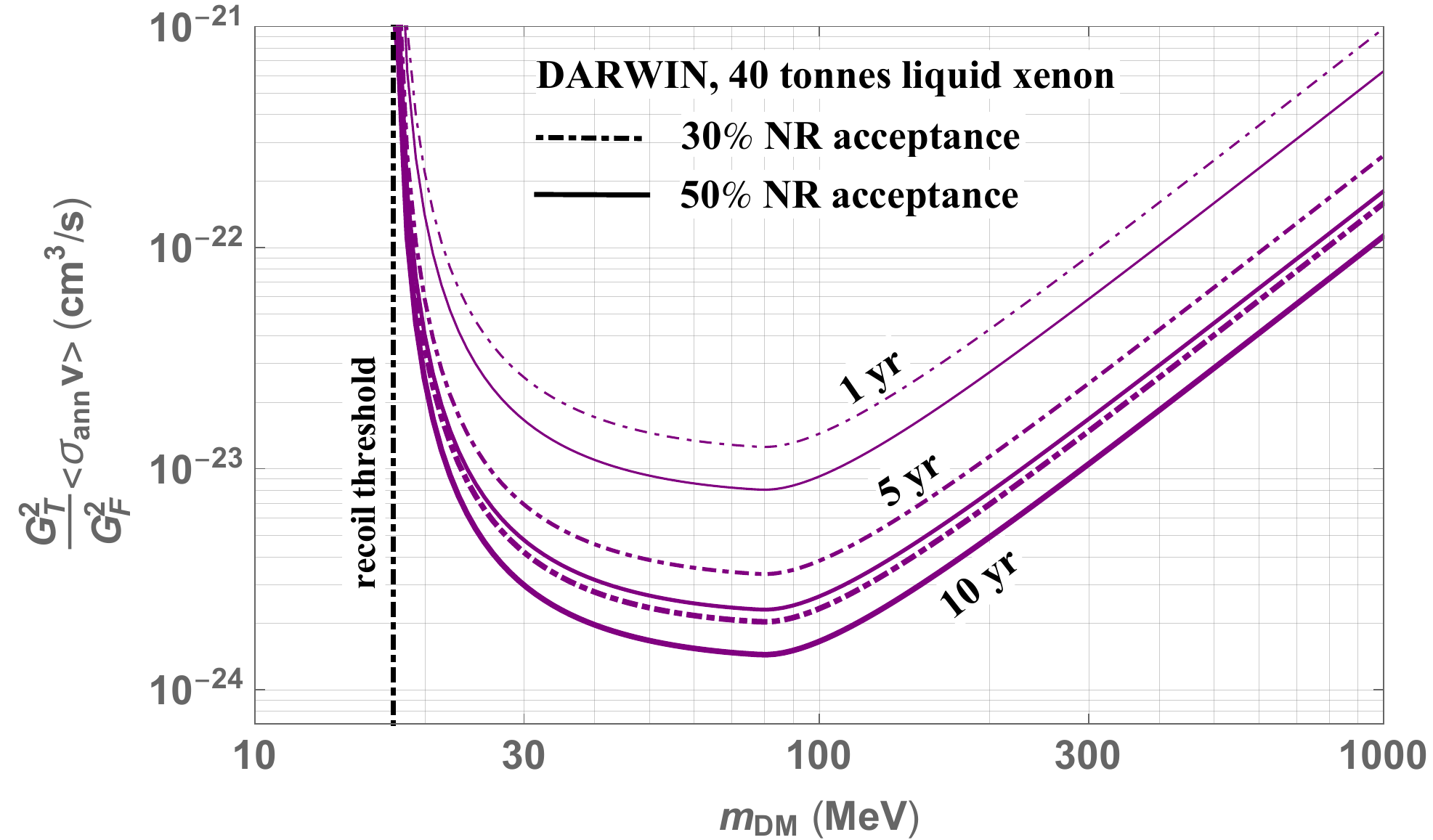} \quad
\includegraphics[width=.48\textwidth]{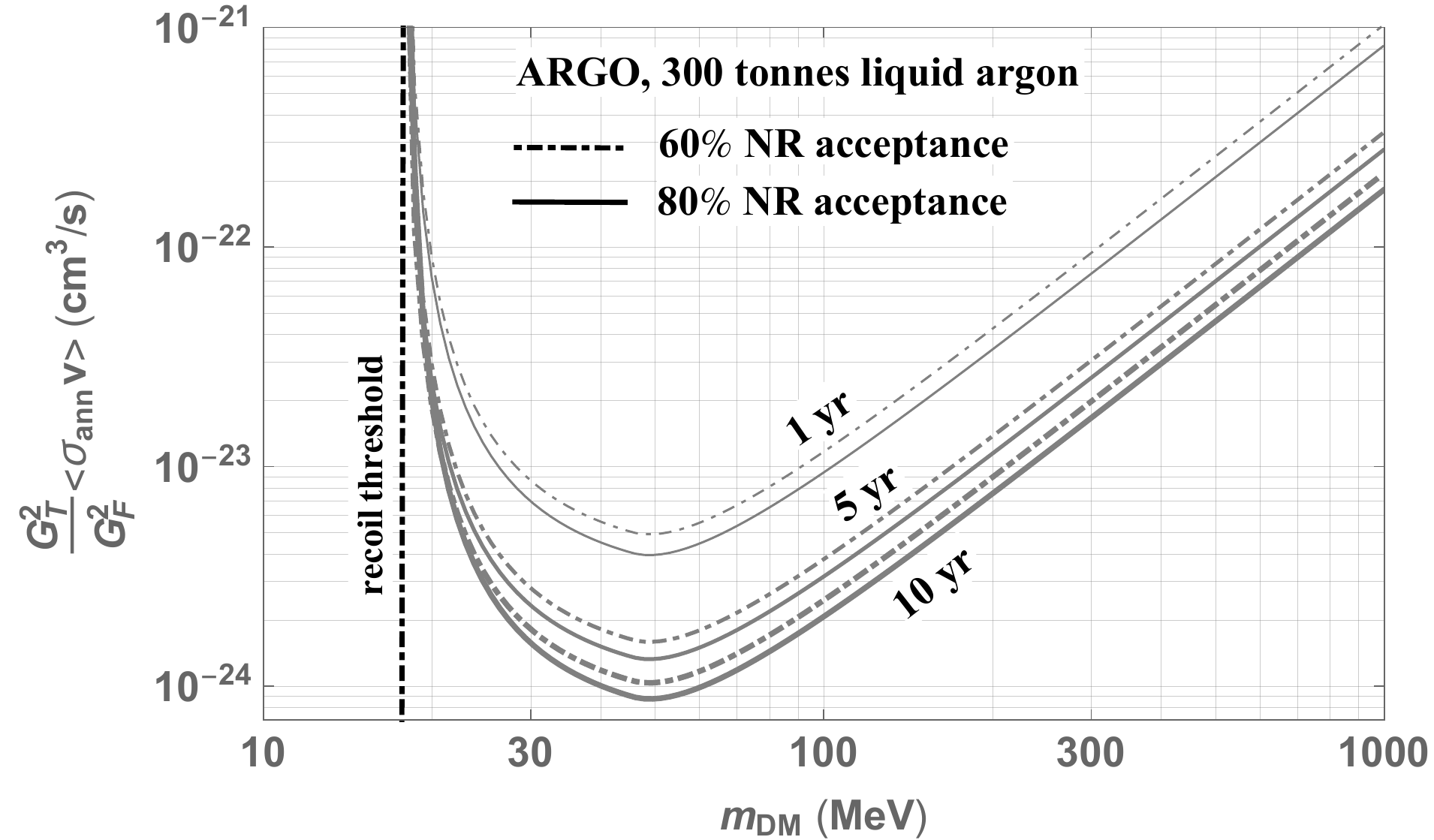}
\caption{
\osn{90}\% \acro{c.l.} sensitivities of the 
liquid xenon \acro{darwin} (\textbf{{\em left}})
and liquid argon \acro{argo} (\textbf{{\em right}}) 
 detectors, 
with fiducial masses of 40 tonnes and 300 tonnes respectively,
to $\sigmaveeann \ \times$  the coupling of the annihilation products to nucleons (normalized to the Fermi constant), as a function of dark matter mass.
An \acro{nfw} halo profile is assumed. 
The sensitivities are shown for 1, 5 and 10 years of exposure, and for nuclear recoil acceptances of 
\osn{30}\% and \osn{50}\% for \acro{darwin}, and 
\osn{60}\% and \osn{80}\% for \acro{argo}.
The electronic recoil rejection is assumed to be at a level that renders background leakage from solar neutrino-electron scattering negligible.
See Sec.~\ref{subsec:limits} for more details.
}
\label{fig:reaches_acceps}
\end{figure*}

\begin{figure*}
\includegraphics[width=.5\textwidth]{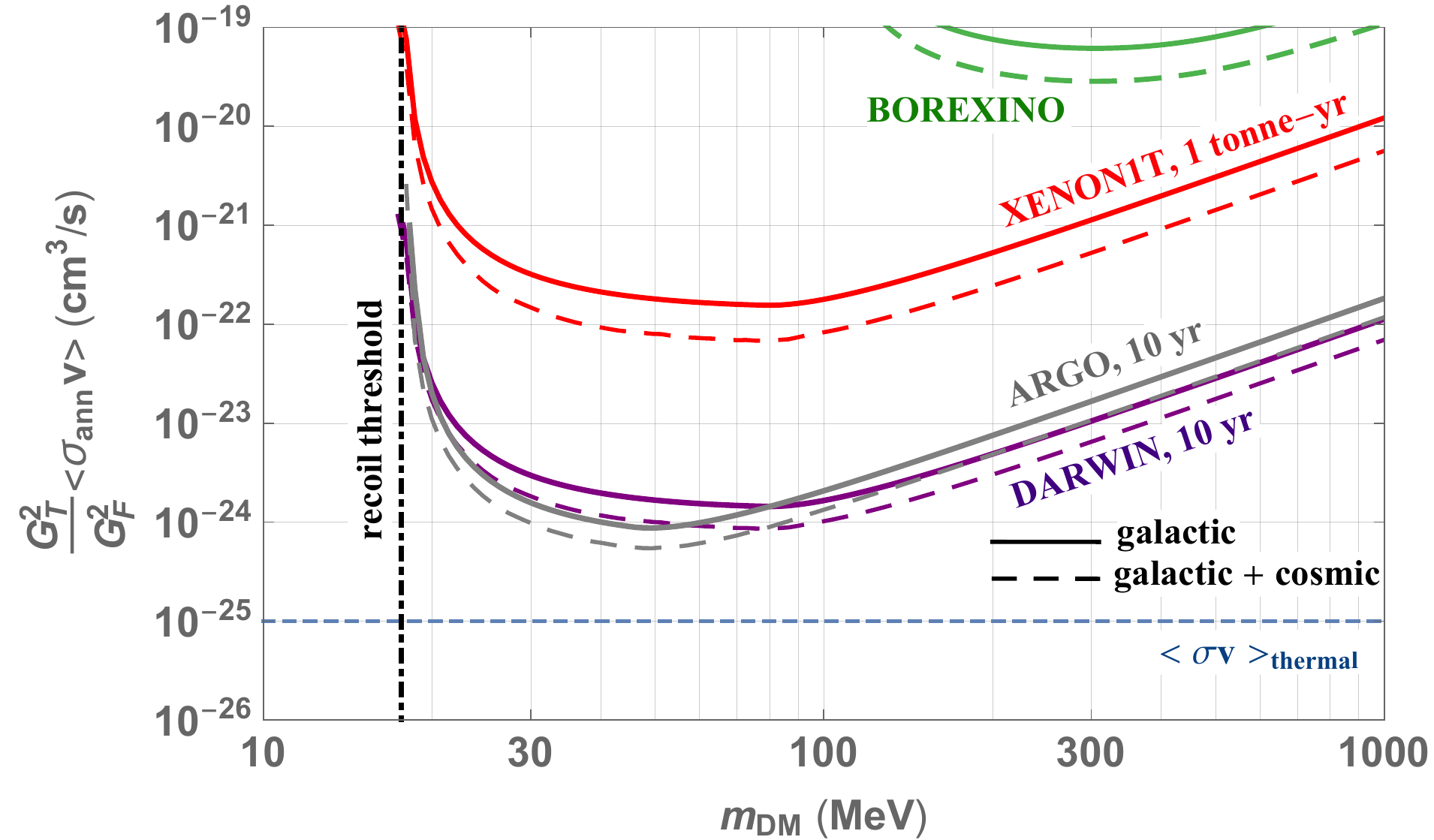} \quad
\includegraphics[width=.47\textwidth]{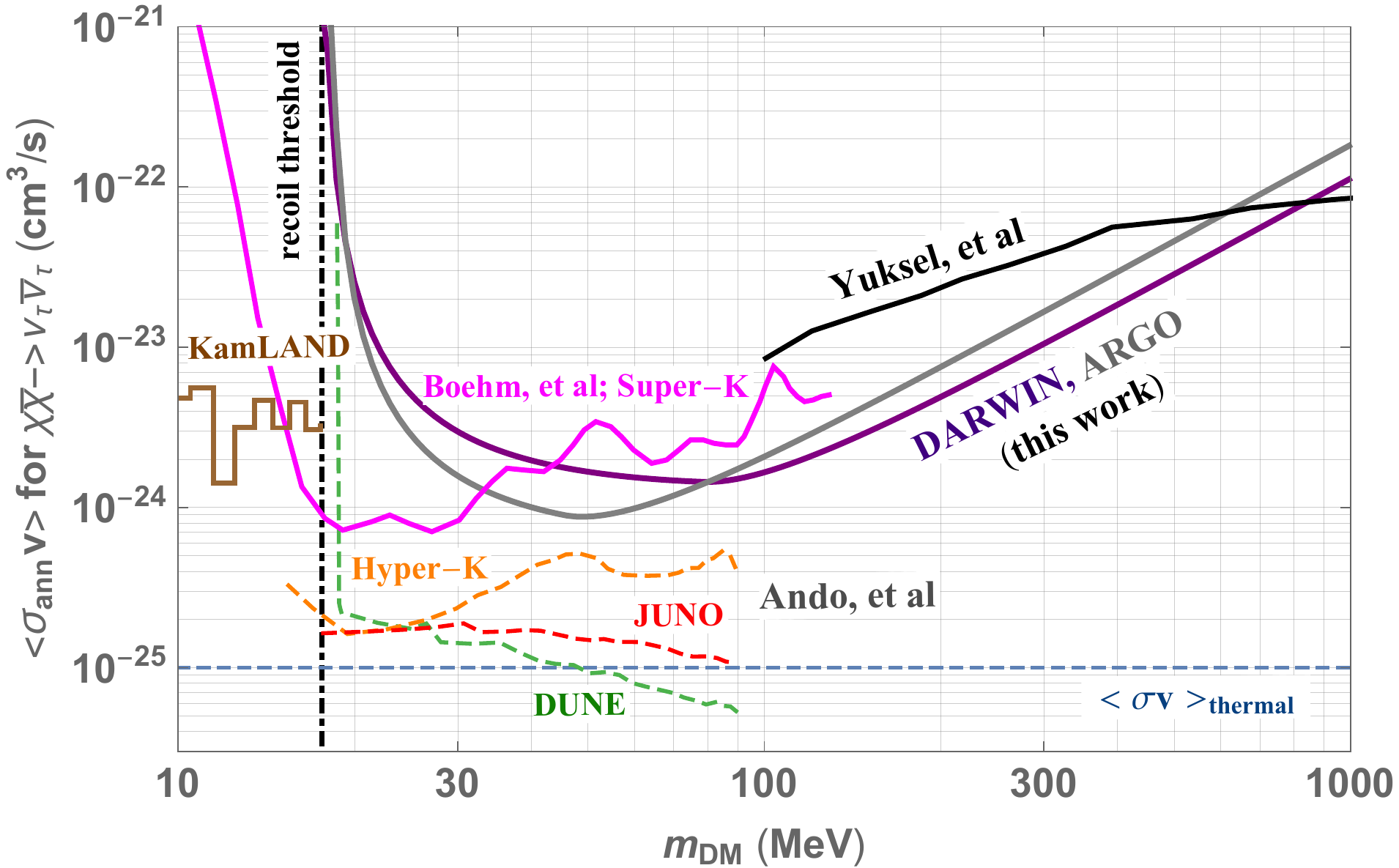}
\caption{
\textbf{{\em Left.}} Same sensitivities as Fig.~\ref{fig:reaches_acceps}, of the detectors 
\acro{darwin} (with 10 years of run-time and \osn{50}\% \acro{nr} acceptance),
\acro{argo} (with 10 years of run-time and \osn{80}\% \acro{nr} acceptance), and
\XoneT~after 1 tonne-year of exposure.
Also shown is the bound from proton recoils in \acro{borexino} (derived in Appendix~\ref{app:borex}).
For all these bounds an \acro{nfw} halo profile is assumed.
These are compared with sensitivities from a possible extragalactic flux, which we compute using the approximate method described in Sec.~\ref{subsec:flux}.
\textbf{{\em Right.}}
Comparison of \acro{darwin} and \acro{argo} sensitivities (as shown in the left panel) to a galactic flux of dark neutrinos from the channel $\chi\bar\chi\ra\nu_\tau\bar\nu_\tau$, with current and future neutrino detectors.
Direct detection would surpass the bound from Super-K relic neutrino searches \cite{1711.05283} and
that derived by Yuksel, {\em et al.} \cite{0707.0196} in the dark matter mass range 35 MeV -- 800 MeV.
If dark neutrinos are produced instead as mass eigenstates in majoron-mediated dark matter annihilations, $\chi\bar\chi\ra\nu_i\bar\nu_i$ ($i$=1,2,3), the bounds from neutrino experiments would be weakened by an $\Oc(10)$ factor while the \acro{darwin} and \acro{argo} sensitivities are not affected.  
See Sec.~\ref{subsec:limits} for more details.
} 
\label{fig:reaches}
\end{figure*}

\subsection{Sensitivities and other constraints}
\label{subsec:limits}

We now obtain the sensitivities for various detector exposures.
In a realistic detector, the rejection of electronic recoil (\acro{er}) backgrounds comes at the cost of nuclear recoil (\acro{nr}) acceptance.
It was determined by \cite{1506.08309} that at \acro{darwin}, where \acro{er}s and \acro{nr}s are distinguished by comparing $S1$ and $S2$ scintillation pulses, it is possible to achieve 
\osn{99}.\osn{98}\% \acro{er} rejection with \osn{30}\%-\osn{50}\% \acro{nr} acceptance,
which is at a level that renders \acro{er} backgrounds negligible.
This study was performed with 
$\ER \in$  [5 keV, 20.5 keV]
to optimize for signal vs background for light \acro{wimp}s, which have rapidly falling recoil spectra.
Our signal spectra fall much more slowly (Fig.~\ref{fig:rates}) and we use 
$\ER \in$  [5 keV, 100 keV], 
nevertheless we adopt the above \acro{nr} acceptances to set our future limits. 
At liquid argon detectors, which use the pulse shape discrimination technique to reject \acro{er} backgrounds, the \acro{nr}s are better distinguished, yielding higher \acro{nr} acceptances. 
We assume that acceptances at the level of \osn{60}\%-\osn{80}\% would be achieved at \acro{argo} \cite{private} with an \acro{er} rejection rate that renders \acro{er} backgrounds negligible.

We next assume that atmospheric and diffuse supernova neutrinos constitute our sole background.
With the above \acro{nr} acceptances (which we denote by $\epsilon_{\rm NR}$) at \acro{darwin} and \acro{argo}, we may safely neglect the leakage of solar neutrino-electron scattering events into our neutrino-nucleus scattering regions.
The number of signal and background events at a detector before and after accounting for the \acro{nr} acceptance is simply related by $S_{\rm acc} = \epsilon_{\rm NR} S$, $B_{\rm acc} = \epsilon_{\rm NR} B$.
Using Poisson statistics, the \osn{90}\% \acro{C.L.} bound is then obtained by solving
\beq
\frac{\Gamma(B_{\rm acc}+1,S_{\rm acc}+B_{\rm acc})}{B_{\rm acc}!} = 0.1~,
\eeq
where $\Gamma$ is the incomplete gamma function.
This allows us to set $\epsilon_{\rm NR}$-dependent bounds on the flux of dark neutrinos and boosted dark matter.

In Fig.~\ref{fig:reaches_acceps} we show the sensitivity of \acro{darwin} and \acro{argo} after exposure times of 1, 5, and 10 years. 
We show this as a function of dark matter mass for the quantity $(\GBT^2/\GF^2) \sigmaveeann$, which collects the unknown parameters in Eq.~\eqref{eq:ratescat}.
The solid and dashed curves indicate \acro{nr} acceptances of 
\osn{50}\% and \osn{30}\% respectively at \acro{darwin}, and 
\osn{80}\% and \osn{60}\% respectively at \acro{argo}.
As expected, due to the atmospheric and diffuse supernova neutrino background,
the sensitivities do not improve linearly with the exposure.

In the left panel of Fig.~\ref{fig:reaches} we show the 10-year \acro{darwin} and \acro{argo} sensitivities with \osn{50}\% and \osn{80}\% \acro{nr} acceptances respectively, along with sensitivities from a possible extragalactic flux obtained by the rescaling procedure described in Sec.~\ref{subsec:flux}.
It must be noted that the true sensitivity lies somewhere between these curves due to the non-monochromatic nature of the extragalactic neutrino flux.
Due to the effect of the maximum $\ER$ imposed by kinematics and our range of integration (see Sec.~\ref{subsec:scatter}), these two detectors would probe small cross sections in complementary $\mdm$ ranges.
We also show with red curves the sensitivity of liquid xenon detectors after 1 tonne-year of exposure.
Recently \XoneT~conducted a search for \acro{wimp}s with this exposure \cite{1805.12562} in nuclear recoil energies $\ER \in$ [4.9 keV, 40.9 keV], with a signal selection efficiency of \osn{85}\% and with an \acro{er}-induced \acro{nr} background of 2 events.
For deriving our sensitivity, we take the range $\ER \in$ [5 keV, 100 keV] to compare with \acro{darwin}, \osn{85}\% signal selection, and no backgrounds.
Lastly, we also show with green curves bounds from measurements of harder-than-solar neutrinos by the liquid scintillator neutrino experiment \acro{borexino} \cite{0911.0548}.
In Appendix \ref{app:borex} we describe
the details of this bound, including the method used for extracting information about dark matter-proton scattering from electron recoil energies.
B\acro{orexino} constrains the dark neutrino flux worse than \XoneT, but constrains boosted dark matter at comparable and stronger levels, as we will see in Sec.~\ref{subsec:boosted}.

In the right panel of Fig.~\ref{fig:reaches} we show our sensitivities to the annihilation cross section of $\chi \bar{\chi} \ra \nu_\tau \bar{\nu}_\tau$ by setting $\GBT = \GF$. 
Strictly speaking, of course, this sensitivity must be multiplied by the branching fraction into this annihilation channel (taken here to be \osn{100}\%).
The dark neutrino flux is assumed to originate from dark matter in the Milky Way alone. 
For \acro{darwin}, we assume \osn{50}\% \acro{NR} acceptance, and 
for \acro{argo}, \osn{80}\% \acro{NR} acceptance.
These are compared with bounds from neutrino experiments.
The brown curve is obtained from a \osn{90}\% \acro{c.l.} limit set by the liquid scintillator detector experiment Kam\acro{LAND} \cite{1105.3516} on extraterrestrial $\bar{\nu}_e$ fluxes in the 8.3--18.3 MeV energy range\footnote{The neutrino flux in this energy range can also be potentially bounded by dark matter direct detection searches that look for electronic recoils in ionization signals from nuclear recoils \cite{1801.10159}; however, the uncertainties are too large for these searches to compete with dedicated neutrino experiments such as Kam\acro{LAND}.}. 
The magenta curve is the \osn{90}\% \acro{c.l.} limit, as recast by \cite{1711.05283}, from a search for diffuse supernova neutrinos by the water \^{C}erenkov detector Super-Kamiokande \cite{1111.5031} using 2853 days of data.
The relevant search channels are inverse beta decay 
($\bar{\nu}_e + p \ra e^+ + n$)
and absorption in oxygen 
($\nu_e/\bar{\nu}_e + ^{16}$O $\ra e^\pm + X$), 
sought in the positron energy range 16 -- 88 MeV, which allows for probing $E_\nu \leq$ 130 MeV \cite{1711.05283}.

The dashed curves are \osn{90}\% \acro{c.l.} sensitivities, as estimated by \cite{1809.00671}, of the future neutrino detectors Hyper-Kamiokande, \acro{dune}, and \acro{juno}.
The search channels at the water \^{C}erenkov detector Hyper-K are the same as Super-K above;
at the liquid argon detector \acro{dune} it is the charged current process $\nu_e/\bar{\nu}_e + ^{40}$Ar $\ra e^\pm + \{X\}$ (where $\{X\}$ are nuclei);
and
at the liquid scintillator detector \acro{juno} these are inverse beta decay and absorption in carbon
($\bar{\nu}_e + ^{12}$C $\ra e^+ + ^{12}$B 
and
$\nu_e + ^{12}$C $\ra e^- + ^{12}$N).
For this study a run-time of 3000 days was assumed, and the backgrounds were estimated by rescaling those of the Super-K analysis above; these sensitivities also depend on the energy resolution assumed for each detector.
Finally, the black curve is the bound (with unspecified \acro{c.l.}), as derived by Yuksel, {\em et al.} \cite{0707.0196}, from atmospheric neutrino measurements by 
Super-K, Fr\'{e}jus, and \acro{amanda}.

For the neutrino experiment bounds, we accounted for the fact that the signal flux for Dirac dark matter is halved compared to self-conjugate dark matter.
Moreover, for the reaches of the future detectors Hyper-K, \acro{dune} and \acro{juno} we rescaled the total fluxes used in \cite{1809.00671} by the method described in Sec.~\ref{subsec:flux} to account for dark matter annihilations in the Milky Way alone.
Finally, all these bounds are shown assuming that neutrino masses are  Majorana in nature, so that $\nu_e$ and $\bar\nu_e$ are not distinguished at neutrino experiments. 
Taking neutrinos to be Dirac would imply rescaling our flux in Eq.~\eqref{eq:flux} by appropriate factors. 
For instance, this flux must be divided by 2 for the Kam\acro{land} bound.
For the bounds at the other neutrino experiments, a more detailed treatment of the flux will be required.

We see in Fig.~\ref{fig:reaches} that \acro{darwin} and \acro{argo} would rival, even outdo, the bounds from Super-K diffuse neutrino searches (Boehm, {\em et al.}) and atmospheric neutrino measurements (Yuksel, {\em et al.}). 
This is because, as mentioned in the Introduction, though large-volume neutrino detectors operate at greater exposures than dark matter detectors, the latter would compensate via the higher event rates of coherent nuclear scattering.

Furthermore, the neutrino detector bounds depend on the fraction of the dark neutrino flux that is electron-flavored. 
In the case of $\chi\bar\chi\ra\nu_\tau\bar\nu_\tau$, the propagation of neutrinos over galactic distances washes out coherent oscillations, so that this fraction is simply the combined electron component of the mass eigenstates produced at the source. 
We calculate this fraction in Appendix~\ref{app:nuflavors}, finding that the $e:\mu:\tau$ flavor ratio upon arrival at earth is 11 : 19 : 20, i.e. the electron fraction is 0.22. 
We have accounted for this reduction of effective flux when presenting our neutrino experiment bounds.
Another interesting case is that of dark matter annihilating directly into neutrino mass eigenstates, $\chi\bar\chi\ra\nu_i\bar\nu_i$, so that no oscillations occur.
In theories where neutrinos acquire masses through lepton number breaking (see \cite{1701.07209} and references therein),
such annihilations are mediated in the $s$-channel by the majoron, the pseudo-scalar associated with the symmetry breaking, which couples to the mass eigenstates in proportion to the neutrino eigenmass $m_{\nu,i}$.\footnote{Obtaining a large annihilation cross section in this scenario is a model-building challenge that is beyond the scope of our work. 
An interesting related possibility is a ``secluded" scenario \cite{0711.4866} where the dark matter annihilates with a large cross section to a pair of on-shell mediators which then each decay dominantly to $\nu_3\bar\nu_3$.
Such neutrinos are no longer monochromatic, yielding non-trivial signatures at direct detection and neutrino experiments.
}
Then the branching fractions into $\nu_i\bar\nu_i$ are proportional to $m^2_{\nu,i}$.
In this case, assuming normal hierarchy of neutrino masses we find that the flavor ratio upon arrival at earth is 3 : 52 : 45 (see Appendix~\ref{app:nuflavors}), so that the electron fraction of the flux is 0.03.
A scenario such as this would therefore weaken the bounds from neutrino experiments shown in Fig.~\ref{fig:reaches} by an order of magnitude, so that direct detection bounds are far stronger.

While we have shown the reaches of future neutrino experiments for energies $\lsim$ 100 MeV, there are no existing studies on these reaches for higher energies.
The exact sensitivities should depend on backgrounds from atmospheric neutrinos, and the systematics and energy resolutions of these detectors in this energy range.
Estimating these is beyond the scope of our work, however, on the strength of the reaches in the sub-100 MeV range, it may be surmised that the reaches in the $>$ 100 MeV energy range may be somewhat stronger than \acro{darwin} and \acro{argo} in most regions for the case of tau-flavored dark neutrinos.
For the case of majoron-mediated dark neutrinos discussed above, dark matter direct detection reaches may still be stronger in most regions.
To make matters even more interesting, the liquid scintillator detector \acro{juno} may see proton recoils from dark neutrinos that are {\em not} electron-flavored, {\em \`{a} la} \acro{borexino} as discussed above.
Therefore, in the event of discovery, an intricate interplay of dark matter and neutrino experiments will be required to discern whether dark neutrinos are produced in flavor or mass eigenstates.

We end this sub-section by noting that the smallest annihilation cross sections that may be probed by direct detection experiments would exceed $10^{-25}~\rm cm^3/s$, the value required for obtaining the observed dark matter relic abundance $(\Omega_\chi h^2 = 0.12$) via thermal freeze-out.
If a dark neutrino flux is detected at \acro{darwin} and \acro{argo}, this would imply that dark matter acquired the observed abundance through a non-standard cosmological history, such as, e.g., late decays of long-lived states into a dark matter population.

\begin{figure}
\includegraphics[width=.48\textwidth]{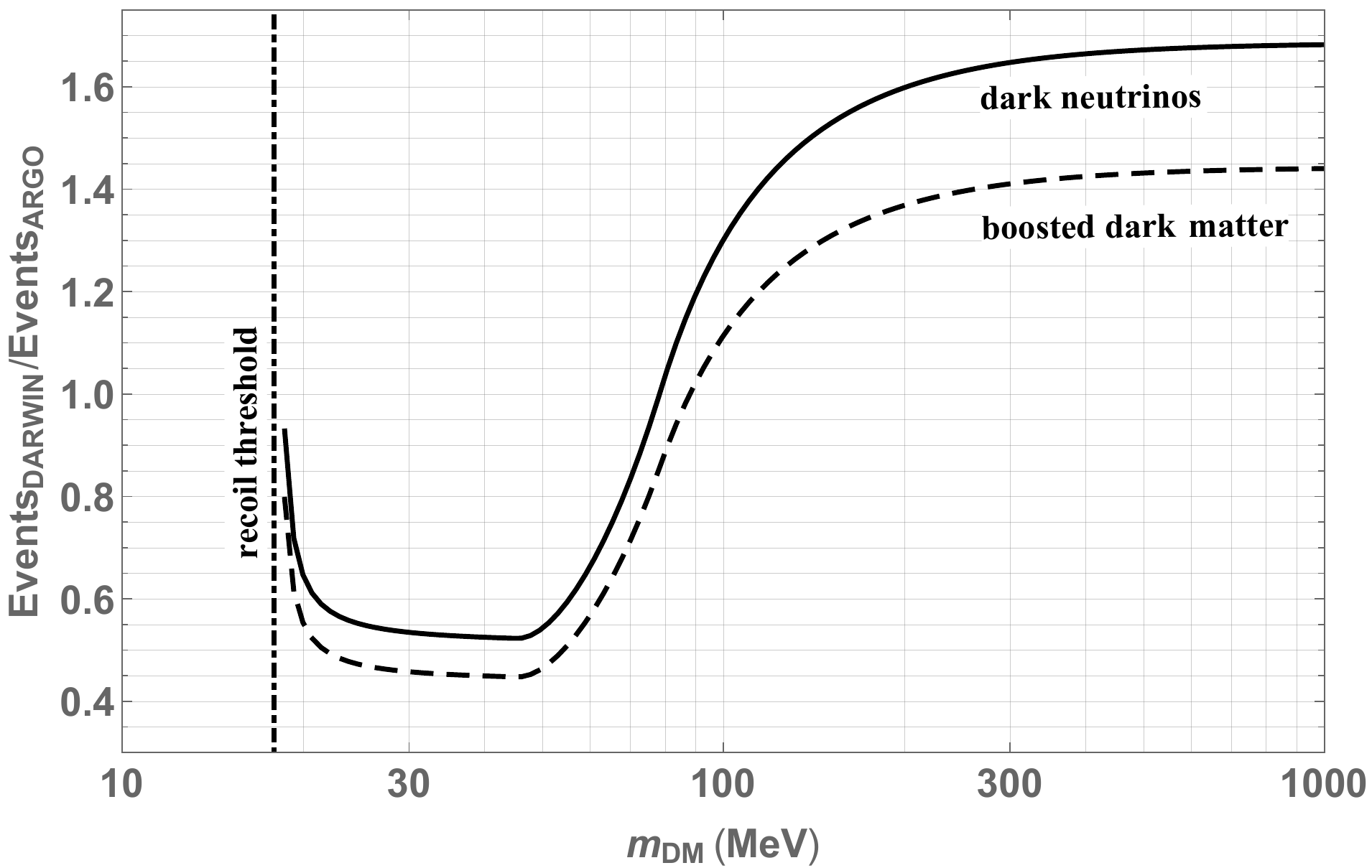} 
\caption{
A simple diagnostic to distinguish between dark neutrinos and boosted dark matter, as described in Sec.~\ref{subsec:telling}.
Shown here is the ratio of background-subtracted events at \acro{darwin} and \acro{argo} after equal run-times, as a function of dark matter mass.
Since neutrinos scatter preferentially on neutrons whereas boosted dark matter scatters equally on neutrons and protons, and since xenon and argon contain different numbers of nucleons, this ratio separates the two cases.}

\label{fig:telling}
\end{figure}

\subsection{Distinguishing between dark neutrinos and boosted dark matter}
\label{subsec:telling}

If a novel flux of coherently scattering particles is discovered at dark matter direct detection searches, it would be of vital importance to identify the species detected.  
As said in the Introduction, dark neutrinos may be distinguished from non-relativistic \acro{wimp}s and boosted dark matter if corroborating, directional signals are obtained at neutrino experiments.
We now show that it is in principle possible to distinguish between dark neutrinos and boosted dark matter as well, given sufficient statistics.

First we note that there is a degeneracy among the unknown parameters $\GBT$, $\sigmaveeann$ and $\mdm$ in the scattering rate, Eq.~\eqref{eq:ratescat}.
If both \acro{darwin} and \acro{argo} see positive signals, then $\sigmaveeann$ may be eliminated in the ratio of integrated rates at either detector, Events$_{\rm DARWIN}$/Events$_{\rm ARGO}$.
Now we recall that $\GBT$ is species-dependent (Eq.~\eqref{eq:GBT}), since neutrinos scatter preferentially on neutrons whereas boosted dark matter scatters democratically on both neutrons and protons, as well as is target-dependent, since xenon and argon nuclei comprise different nucleon populations. 
Thus Events$_{\rm DARWIN}$/Events$_{\rm ARGO}$ would distinguish between dark neutrinos and boosted dark matter for a given projectile energy (= dark matter mass).
We have plotted this quantity (with the backgrounds subtracted) for the two species in Fig.~\ref{fig:telling}, as a function of $\mdm$.
The scattering rates have been integrated over the $\ER$ ranges mentioned in Sec.~\ref{subsec:scatter}, and equal run-times at both experiments are assumed.
The remaining degeneracy, that of dark matter mass, may be broken by inspecting the recoil spectrum at high energies, as the kinematic endpoint is set by this mass (Eq.~\eqref{eq:ermaxkinem}).

We point out another virtue in simultaneously inspecting recoil spectra at \acro{darwin} and \acro{argo}.
If a dark neutrino flux is detected with a small number of signal events, the background model used for the atmospheric and diffuse supernova fluxes would come into question, in which case such an inspection would prove valuable.
A detailed statistical study characterizing signals at various detector materials is an exercise we reserve for future exploration.

\section{Interpretations}
\label{sec:interpret}

In this section, we interpret our model-independent results in the previous section in terms of a model of dark matter that interacts with the \acro{sm} through a neutrino portal, and models of nucleophilic boosted dark matter.

\subsection{Neutrino portal dark matter}
\label{subsec:nuportal}

\begin{figure}
\includegraphics[width=.48\textwidth]{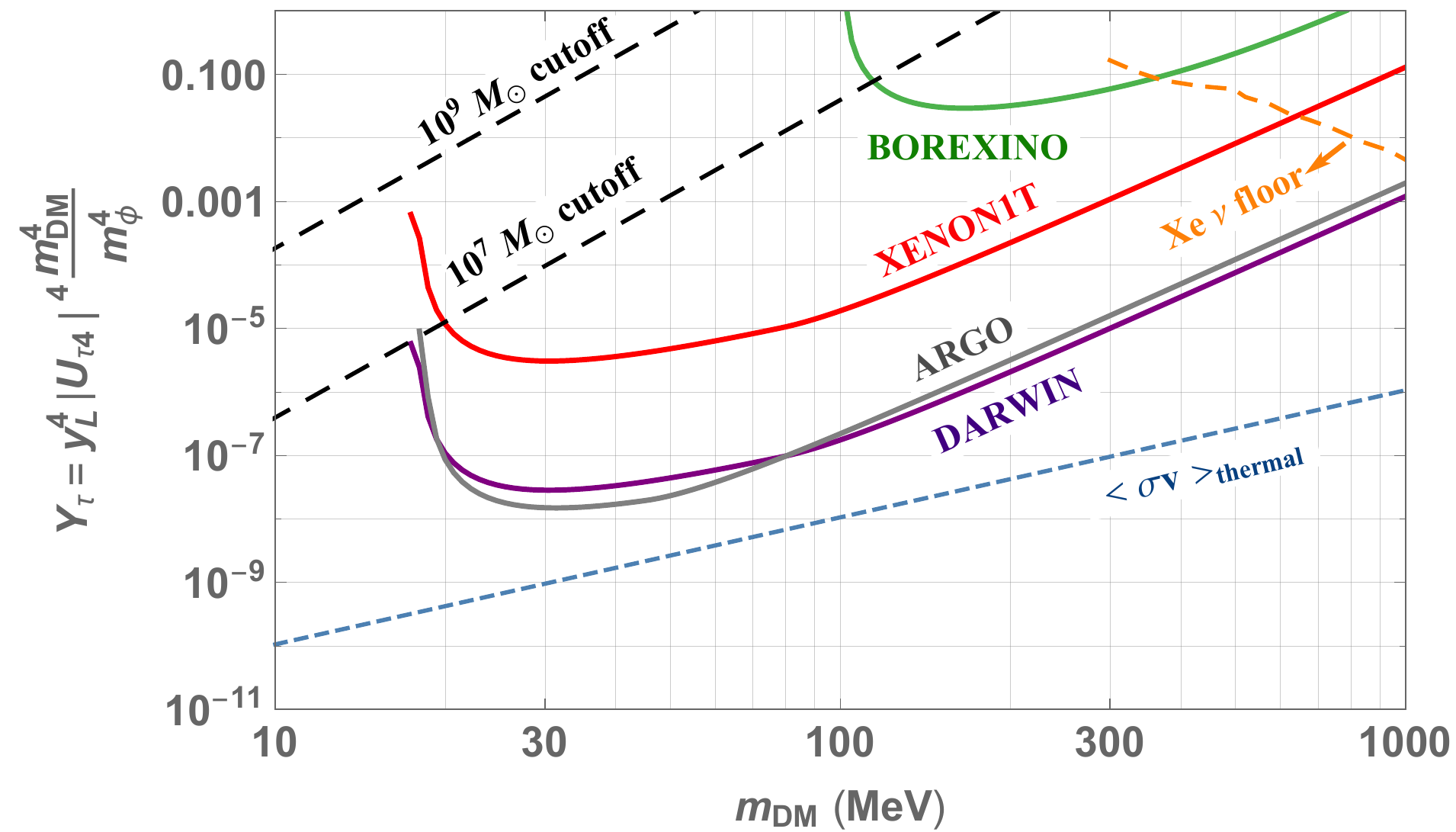} 
\caption{
Interpretation of our results in the left panel of Fig.~\ref{fig:reaches} in terms of the neutrino portal dark matter model described in Sec.~\ref{subsec:nuportal}.
Bounds are shown in the plane of the effective coupling $Y_\tau$ (Eq.~\eqref{eq:Ytau}) vs dark matter mass, with regions above the curves excluded.
Dashed black curves indicate parameters that result in suppression of the growth of structures below a mass cutoff of $10^9$ and $10^7$ solar masses, thus explaining the ``missing satellites" problem.
These regions are seen to be extensively probed by direct detection.
The dashed orange curve denotes parameters that correspond to a dark matter-nucleon scattering cross section associated with the standard ``neutrino floor" in xenon.
The dashed blue line corresponds to a thermal annihilation cross section = $10^{-25}$ cm$^3$/s.
} 
\label{fig:interpretnuportal}
\end{figure}

Dark matter annihilation into neutrinos is a generic feature of simple neutrino portal dark matter models~\cite{Bertoni:2014mva,Batell:2017rol,Batell:2017cmf}. 
In such a setup, dark matter couples to the \acro{sm} through the operator $HL$, where $H$ is the \acro{SM} Higgs doublet and $L$ is a lepton doublet containing the neutrino $\nu_L$. 
Taking dark matter to be a Dirac fermion $\chi$,
stabilizing it requires charging it under, e.g., a $\mathbb Z_2$ symmetry, and therefore, coupling it to $HL$ requires introducing a charged complex scalar $\phi$ heavier\footnote{It is also possible to have $\phi$ lighter than $\chi$, so that $\phi$ is now the dark matter and $\chi$ the mediator. 
However, the dark matter annihilation proceeds in the $p$-wave, so that annihilation signals today are absent.
} 
than $\chi$. 
Then the dark matter couples to neutrinos via the effective operator $\phi\bar\chi HL/\Lambda\to (v/\Lambda)\phi\bar\chi\nu$ under electroweak symmetry breaking, where $\langle H\rangle=v=174~\rm GeV$ and $\Lambda$ is some high scale. 
In the \acro{uv} completion, these interactions can arise from a Dirac sterile neutrino $N$ that couples to both the \acro{SM} and ``dark'' sectors. 
The relevant interactions are contained in
\beq
\begin{aligned}
&-\delta{\cal L} = m_N\bar NN+\mdm\bar\chi\chi+m_\phi^2\left|\phi\right|^2
\\
&+\left[\lambda_\ell\bar L_\ell i\tau^2H^\ast N_R+\phi\bar\chi\left(y_LN_L+y_RN_R\right)+{\rm h.c.}\right].
\end{aligned}
\eeq
Note that we can assign lepton number to $N$ and $\chi$ or $\phi$ such that it remains conserved in $\delta {\cal L}$. 
Here $\ell$ is a flavor index. If we assume that the coupling of one flavor is dominant, then the linear combination $\nu\equiv\sqrt{1-|U_{\ell 4}|^2}\nu_{L\ell}-U_{\ell 4} N_L$ forms a light (and at this level, massless) neutrino that couples to dark matter through the interaction
\beq
-{\cal L}_{\rm int} = y_L U_{\ell 4} \phi\bar\chi\nu+{\rm h.c.}
\eeq
The orthogonal combination pairs up with $N_R$ to form a heavy Dirac neutrino with mass $M=\sqrt{\lambda_\ell^2v^2+m_N^2}$. 
The active-sterile mixing angle is $U_{\ell 4}=\lambda_\ell v/M$. 
Note that this mixing angle need not be tiny for the mostly-active neutrino to be extremely light.\footnote{Simple extensions to the model can confer finite masses to the light neutrinos while keeping the mixing angle relatively large; see~\cite{Batell:2017cmf} for details.}

We assume that $\mdm < m_N/2$ so that it can only annihilate to mostly-active neutrinos, $\chi\bar\chi\to\nu\bar\nu$. 
In this case, the heavy neutrino decay mode $N \ra \chi\bar\chi\nu$ is fully invisible, implying that 
the limits on the mixing angle $U_{\ell 4}$ can be relatively weak, especially in the case that the dominant mixing is with the $\tau$ neutrino. 
Focusing on this possibility, the dark matter annihilation cross section can be written
\beq
\sigmaveeann = \frac{Y_\tau}{32\pi\mdm^2}\left(1+\frac{\mdm^2}{m_\phi^2}\right)^{-2}~,
\eeq
with
\beq
Y_\tau \equiv y_L^4 |U_{\tau 4}|^4 \frac{\mdm^4}{m_\phi^4}~.
\label{eq:Ytau}
\eeq
Comprehensive bounds on this effective coupling have been derived in Ref.~\cite{Batell:2017cmf}.

A large mixing angle, or equivalently large effective coupling $Y_\tau$, can have interesting implications for structure formation. 
The interaction of the dark matter and the light neutrinos can be large enough to delay the onset of the growth of structure~\cite{Aarssen:2012fx,Shoemaker:2013tda,Ko:2014bka,Archidiacono:2014nda,Cherry:2014xra,Bertoni:2014mva}, suppressing the number of structures with mass below some critical scale, $M_{\rm cut}$, determined by $Y_\tau$. 
This has been cited as a potential solution to the so-called ``missing satellites problem''. 
Despite the large coupling needed to affect structure formation, probing this region of parameter space in other physical systems is difficult, largely due to the challenge of making and detecting $\tau$ neutrinos.

\begin{figure*}
\includegraphics[width=.45\textwidth]{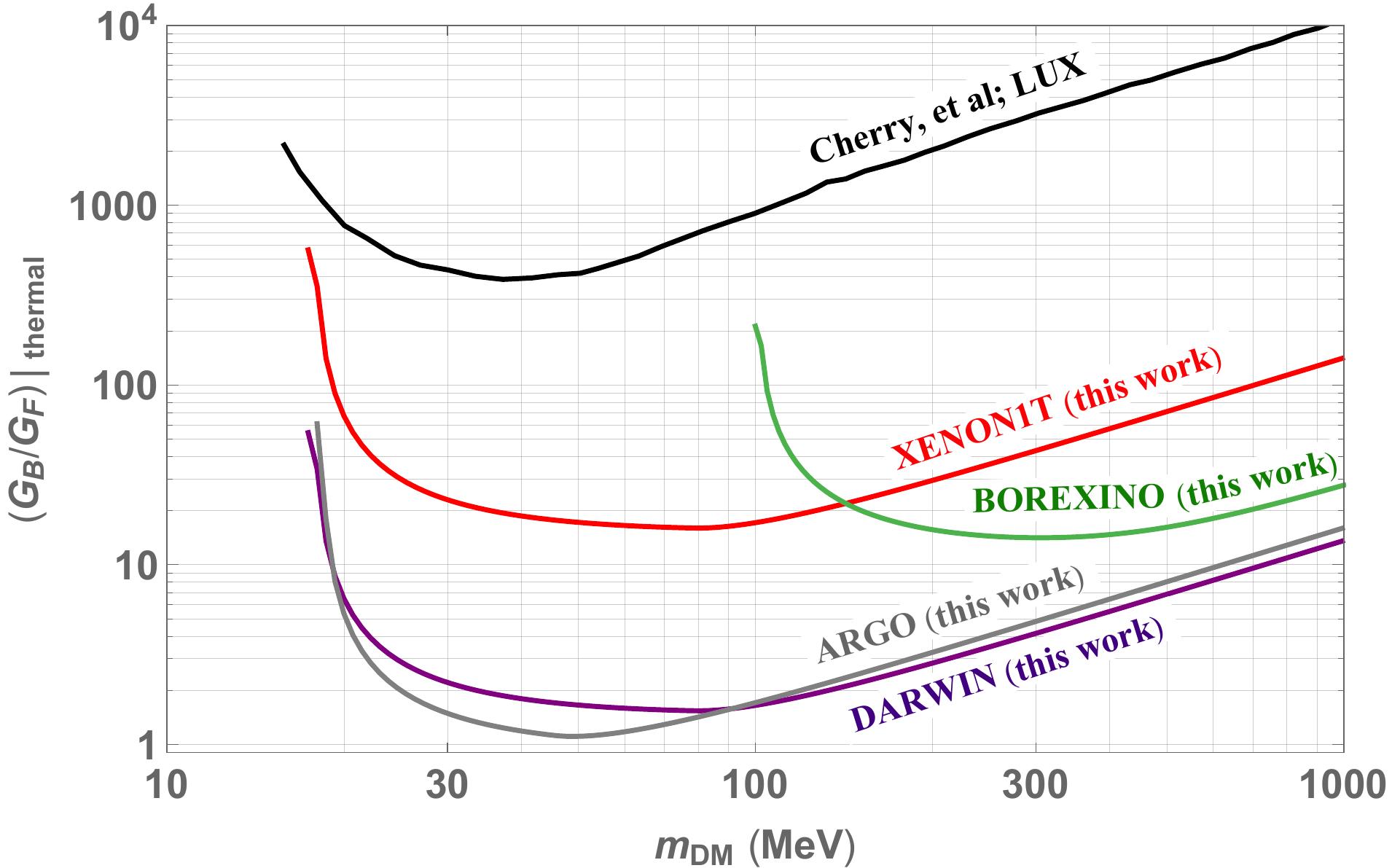} \quad
\includegraphics[width=.45\textwidth]{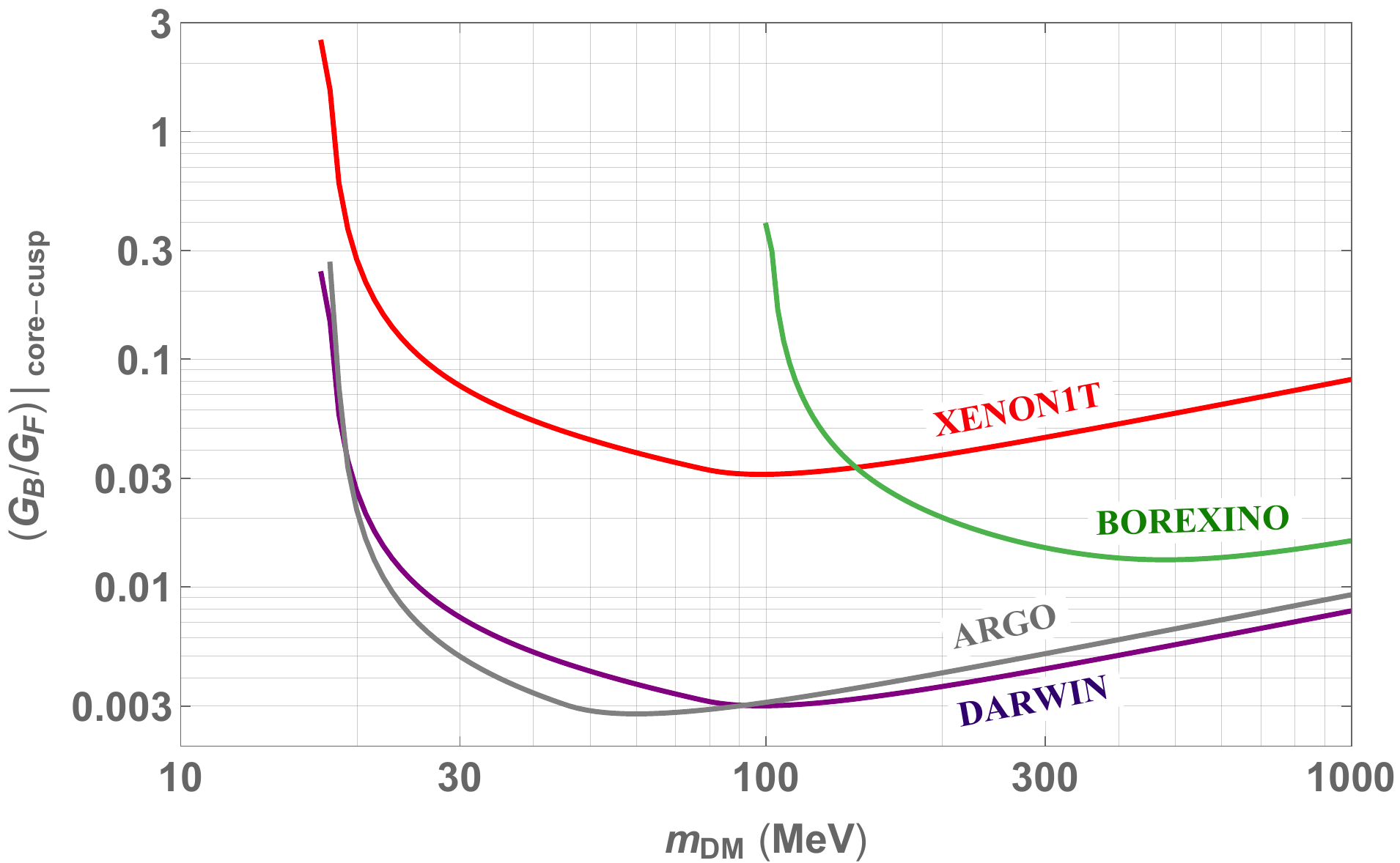}
\caption{
Interpretation of our results in the left panel of Fig.~\ref{fig:reaches} in terms of the boosted dark matter models described in Sec.~\ref{subsec:boosted}.
\textbf{\em Left.} Bounds as a function of dark matter mass on the baryonic coupling $\GB$ (normalized to the Fermi constant $\GF$) such that the boosted dark matter flux is obtained for a thermal annihilation cross section = $10^{-25}$ cm$^3$/s.
The bounds obtained by Cherry, {\em et al.} \cite{1501.03166} used \acro{lux} data.
The erstwhile bound from \acro{borexino} proton recoils (derived in Appendix \ref{app:borex})
and the current bound from \XoneT~are seen to be stronger.
D\acro{arwin} and \acro{argo} can be seen to probe $\GB$ all the way down to the weak coupling $\GF$ in the $\mdm$ range $\sim$ 25--100 MeV.
\textbf{\em Right.} Addressing the core-cusp problem with boosted dark matter detection. Shown are values of $\GB/\GF$ ruled out by \acro{borexino} and \XoneT, and those to be probed by \acro{darwin} and \acro{argo}, for an annihilation cross section (Eq.~\eqref{eq:xsanncorecusp}) that would ensure that dark matter is depleted from halo cores.
} 
\label{fig:interpretboosted}
\end{figure*}

In Fig.~\ref{fig:interpretnuportal}, we show  as a function of $\mdm$ the sensitivities of \acro{darwin} and \acro{argo} to the effective coupling $Y_\tau$, which controls dark matter annihilation.
We assume 10 year exposures and the \acro{nr} acceptances used in Fig.~\ref{fig:reaches}.
Also shown are the bounds from \XoneT~and \acro{borexino} as in the left panel of Fig.~\ref{fig:reaches}.
Only the galactic flux is included.
In addition, we show with dashed black lines the regions of parameter space that result in $M_{\rm cut}=10^7M_\odot$ and $10^9M_\odot$.
We find that direct detection of dark neutrinos could extensively probe these regions.
For reference, we also show the value of $Y_\tau$ that yields a thermal annihilation cross section of $\sigmaveeann= 10^{-25}~\rm cm^3/s$.

As discussed in the Introduction, in this model dark matter develops an effective loop-induced coupling to the $Z$ boson, through which it could scatter with nuclei at direct detection experiments.  
In Fig.~\ref{fig:interpretnuportal} we show with a dashed orange curve values of the coupling $Y$ that correspond to this scattering cross section at the xenon neutrino floor, as done in \cite{Batell:2017cmf}.
In regions above this curve, an interesting possibility may occur: direct detection experiments may register events from coherent nuclear scattering of both the local, non-relativistic dark matter and the relativistic dark neutrino fluxes.
However these regions are already disfavored by the atmospheric neutrino bounds of Yuksel, {\em et al.} (see Fig.~\ref{fig:reaches}).

Finally, we point to the existence of other dark matter models that would result in monochromatic dark neutrinos. 
For example, such dark matter could be 
supersymmetric sneutrinos \cite{Arina:2015zoa} or 
an adjoint fermion in a hidden confining gauge group \cite{Falkowski:2009yz}.
For variations on the neutrino portal model presented in this section, with all possible spins of the dark matter and the mediator, see \cite{1711.05283}.

\subsection{Boosted dark matter}
\label{subsec:boosted}

We now consider the possibility that the annihilation products of dark matter are a second, subdominant component of dark matter that is nucleophilic.
This is realized when the boosted dark matter is a ``baryonic neutrino" whose interactions with baryons are mediated by a gauged $U(1)_{\rm B}$ vector \cite{1103.3261,1311.5764,1711.04531}, or when it is a fermion whose interactions with \acro{SM} fermions are mediated by a scalar that mixes with the Higgs boson \cite{1501.03166}.
The effective coupling to nucleons is then $\GB = g_{\rm x} g_{\rm q}/m^2_{\rm MED}$, where $g_{\rm x}$ and $g_{\rm q}$ are the mediator couplings with the boosted dark matter and quarks respectively, and $m_{\rm MED}$ is the mediator mass.
It has been determined that in both these models, $\GB \lsim 7000 \ \GF$ is experimentally viable \cite{1501.03166}.
We take the boosted dark matter's mass $\ll \mdm$, and this implies that its population can contribute to the energy density of radiation in the early universe, constrained by measurements of $N_{\rm eff}$; however, this contribution is within observed limits \cite{1501.03166}. 

Using Eq.~\eqref{eq:GBT} and the results in Fig.~\ref{fig:reaches}, we may obtain bounds on $\GB$ for a flux corresponding to a thermal\footnote{Since for thermal annihilation cross sections the detectable values of $\GB$ turn out to be $> \GF$, one may reasonably assume that the boosted annihilation products, by virtue of scattering with nucleons, keep dark matter in thermal equilibrium in the early universe until freeze-out.} annihilation cross section $\sigmaveeann_{\rm thermal} = 10^{-25}$ cm$^3$/s; these are shown in the left panel of Fig.~\ref{fig:interpretboosted} as constraints on the ratio $\GB/\GF$ as a function of dark matter mass.
Here we have neglected the attenuation of flux due to passage through the earth, since the couplings being probed are too small for this effect to be important.

The \acro{borexino} bounds, which were the weakest in the case of the dark neutrino flux, now outdo \XoneT~for dark matter masses $\geq$ 200 MeV.
This is because the relative size of the cross sections for scattering with protons vs xenon is much smaller for neutrinos than for boosted dark matter; see Eqs.~\eqref{eq:xsscat} and \eqref{eq:GBT}.
The \acro{borexino} and \XoneT~bounds on $\GB$ are also 2 or 3 orders of magnitude stronger than those placed by \cite{1501.03166} using a 0.027 tonne-year \acro{LUX} dataset.
(We have adapted Ref.~\cite{1501.03166}'s bound on $\sigmaveeann$ for $\GB/\GF = 500$ to present the bound on $(\GB/\GF)|_{\rm thermal}$.)

We see that \acro{darwin} and \acro{argo} would improve the sensitivity to $\GB$ by a factor of 10, probing all the way down to the weak coupling size 1--2 $\times \ \GF$ for dark matter masses $\sim$ 25--100 MeV.
This is one of the main results of our paper.

\textbf{\em The core-cusp problem.}
While probing thermal freeze-out is of inherent interest, another worthwhile target for direct detection is to probe scenarios that address the small scale structure of the universe. 
In Sec.~\ref{subsec:nuportal}, we explored how direct detection signals of dark neutrinos could help address the missing satellites problem. 
Now we show briefly that direct detection signals could tackle the so-called core-cusp problem as well.

From \acro{N}-body simulations of non-interacting cold dark matter, we expect its density to rise steeply near the centers of galactic halos, but observations indicate flat density profiles in these regions. 
Reference~\cite{Kaplinghat:2000vt} posited that if dark matter annihilated today with large rates, it would be especially depopulated in overdense regions, and uniform halo cores would result.  
The requisite $s$-wave annihilation cross section is far above the thermal cross section for our dark matter masses of interest:
\beq
\sigmaveeann_{\rm core-cusp} = 3 \times 10^{-20} \ {\rm cm}^3/{\rm s} \left(\frac{\mdm}{100 \ {\rm MeV}}\right)~.
\label{eq:xsanncorecusp}
\eeq
The annihilation products must not yield photons, since such large $\gamma$ fluxes are well excluded. 
Nor could the annihilation products be neutrinos, though they are the hardest \acro{sm} states to detect, as this too is excluded (see Fig.~\ref{fig:reaches}). 
But if the annihilation products are even harder to detect than neutrinos, such as when they are boosted dark matter with tiny couplings to the \acro{sm}, then a large flux of them could have gone unnoticed, keeping this solution to the core-cusp problem alive. 
In the right panel of Fig.~\ref{fig:interpretboosted} we show the values of the coupling $\GB$ (relative to $\GF$) that are excluded by \XoneT~and \acro{borexino} proton recoils, and probe-able by \acro{darwin} and \acro{argo}, if the boosted dark matter flux corresponded to the annihilation cross section in Eq.~\eqref{eq:xsanncorecusp}. 
We see that in this scenario \acro{darwin} and \acro{argo} would be sensitive to couplings $\sim 10^{-3} - 10^{-2} \times \ \GF$ for dark matter masses $\sim$ 20--1000 MeV. 

\section{Conclusions and discussion}
\label{sec:concs}

In this paper we explored the sensitivity of future noble liquid-based direct detection experiments to the flux of particles produced in the annihilation of dark matter in the sky. 
Our main results are summarized in Figs.~\ref{fig:reaches}, \ref{fig:interpretnuportal}, and \ref{fig:interpretboosted}.
The particle species we considered are neutrinos, which we call ``dark neutrinos", and a second component of dark matter, now commonly known as boosted dark matter, that is nucleophilic.
We derived the reaches of a 40-tonne xenon-based detector, \acro{darwin}, and a 300-tonne argon-based detector, \acro{argo}, after accounting for imminent backgrounds from atmospheric and diffuse supernova neutrinos, as well as realistic nuclear recoil acceptances.
To the best of our knowledge, this is the first work to study a next-to-next generation liquid argon detector for a concrete theoretical scenario. 
Liquid argon technology exploits the pulse shape discrimination technique, which is not possible in liquid xenon, and this enables superior electron recoil rejection.
We have shown that by virtue of this, \acro{argo} would set limits on dark matter parameters that are closely comparable and complementary to \acro{darwin}.
Due to their differing technologies, thresholds and energy ranges of interest, a positive signal at both \acro{darwin} and \acro{argo} would help to mitigate uncertainties in the modeling of backgrounds, especially if signal events are few in number.
It would also help to characterize the particle species detected, e.g., neutrinos vs boosted dark matter.

We found that after 10 years of exposure these experiments would probe dark neutrino fluxes beyond the reach of neutrino experiments.
There are two main reasons for this: (i) while neutrino detectors admit larger fluxes and sustain greater exposures, direct detection experiments benefit from high event rates due to coherent nuclear scattering, (ii) while search channels at most neutrino experiments such as Super-Kamiokande rely on the $\nu_e$ and/or $\bar\nu_e$ fraction of the dark neutrino flux, which may be as small as $\Oc(10^{-2})$ depending on the underlying dark matter model, the search channel of coherent nuclear scattering at direct detection is agnostic to neutrino flavor and self-conjugation.

We also derived current bounds on the boosted dark matter flux, which are set in complementary dark matter mass ranges by 1 tonne-year of data from \XoneT~and by a limit on proton recoils at \acro{borexino}.
For a thermal annihilation cross section, these bounds limit the effective baryonic coupling of the boosted dark matter to $\GB \lsim (10-100) \times$ the Fermi constant $\GF$; in the future, \acro{darwin} and \acro{argo} would push this down to $\GB \lsim (1-10) \times \GF$. 
For an annihilation cross section corresponding to that which is required to deplete dark matter from halo cores, and thus solve the core-cusp problem, \acro{darwin} and \acro{argo} would probe couplings down to $\GB \sim 10^{-3}-10^{-2} \times \GF$.

There are several related avenues of exploration that we had not entered in this work.
While we had assumed that our signals were sourced by dark matter annihilations in free space, similar monochromatic fluxes may also originate in annihilations in the sun, {\em \`{a} la} Refs.~\cite{Barger:2007hj} and \cite{1609.04876}.
Non-monochromatic neutrinos may be sourced by ``cascade annihilations", i.e. annihilations of dark matter into mediators or \acro{sm} states that may subsequently decay to neutrinos in the energy range of interest for direct detection.
While we had implicitly assumed a symmetric population of dark matter, an interpretation of our flux limits in terms of an asymmetric population is possible, especially in the case of boosted dark matter, where our limits on the (symmetric) annihilation cross section could be $< 10^{-25}$ cm$^3$/s -- see \cite{Bell:2014xta} for more details.
In signals appear in more than one direct detection experiment, a halo-independent analysis for relativistic fluxes may be undertaken, such as in \cite{1501.03166}.
An exciting possibility is the double signal briefly mentioned in Sec.~\ref{subsec:nuportal}, comprising of the scattering of both local, non-relativistic dark matter, and relativistic particles produced in dark matter annihilations.
While the parametric regions for this to transpire in the simple neutrino portal model we had explored are already excluded, there may be other theories where this is still viable, especially in models of boosted dark matter.

Goodman and Witten \cite{Goodman:1984dc} originally proposed dark matter direct detection following the proposal of Drukier and Stodolsky \cite{Drukier:1983gj} for detecting neutrinos -- from solar, atmospheric, terrestrial, supernova, reactor, and spallation sources -- in coherent elastic neutral current scattering. 
Should the ``ultimate detectors" discover dark matter by catching neutrinos sourced by dark matter, the revolutionary moment would draw a decades-long search program to a satisfying close.

\section*{Acknowledgments} 
 
This work has benefited from conversations with
Joe Bramante,
Pietro Giampa,
David Morrissey,
and
Tien-Tien Yu.
It is supported by the Natural
Sciences and Engineering Research Council of Canada (\acro{nserc}).
T\acro{RIUMF} receives federal funding
via a contribution agreement with the National Research Council Canada.

\appendix

\section{Flavor components after neutrino propagation}
\label{app:nuflavors}

While coherent nuclear scattering at direct detection searches is blind to neutrino flavor, other channels, used in neutrino experiments, are sensitive to the flavor component being detected.
These components depend both on the distribution of flavors in the neutrino states produced at the source, and neutrino mixing parameters.
In this appendix we calculate these components for the cases of dark matter annihilations into either flavor or mass eigenstates, which we use in Sec.~\ref{sec:reaches} to present bounds from neutrino experiments.

We begin with the case of $\chi\bar\chi \ra \nu_\alpha\bar\nu_\alpha$, where $\alpha$ is a flavor index.
Since these dark neutrinos travel galactic distances before arriving at detectors, the effect of coherent oscillations is negligible, so that the probability of detecting flavor $\beta$ is given simply by
\beq
P_{\alpha\beta} = \sum_{i = 1}^3 |U_{\beta i}|^2|U_{\alpha i}|^2~,
\label{eq:probconvert}
\eeq
where $U$ is the \acro{PMNS} matrix. 
It is usually parameterized as 
\beq
\nn
\begin{bmatrix} 
1 & 0 & 0 \\
0 & c_{23} & s_{23}  \\
0 & -s_{23} & c_{23}
\end{bmatrix}
\begin{bmatrix} 
c_{13} & 0 & s_{13}e^{i\delta_{\rm CP}} \\
0 & 1 & 0  \\
-s_{13}e^{-i\delta_{\rm CP}} & 0 & c_{13}
\end{bmatrix}
\begin{bmatrix} 
c_{12} & s_{12} & 0 \\
-s_{12} & c_{12} & 0  \\
0 & 0 & 1
\end{bmatrix}~,
\eeq
where $c_{ab}$ and $s_{ab}$ denote $\cos\theta_{ab}$ and $\sin\theta_{ab}$ respectively.
The most recent best-fit values \cite{1811.05487,nufit} (without the error bars) are
\bea
\nn \theta_{12} =  33.62^\circ, \ \theta_{23} =  47.2^\circ, \ \theta_{13} =  8.54^\circ, \ \delta_{\rm CP} = 234^\circ~,
\nn
\eea
from which we obtain using Eq.~\eqref{eq:probconvert} the following conversion probabilities:
\beq
\begin{bmatrix} 
P_{ee} & P_{e\mu} & P_{e\tau}  \\
P_{\mu e} & P_{\mu\mu} & P_{\mu\tau}  \\
P_{\tau e} & P_{\tau \mu} & P_{\tau \tau}
\end{bmatrix} =
\begin{bmatrix} 
0.55 & 0.23 & 0.22  \\
0.23 & 0.39 & 0.38  \\
0.22 & 0.38 & 0.40
\end{bmatrix}~.
\eeq

Next we turn to the case of $\chi\bar\chi \ra \nu_i\bar\nu_i$, where $i$ runs over mass eigenstates.
Inspired by majoron-mediated models, we assume that the branching fractions $\propto$ $m^2_{\nu,i}$.
As these neutrinos do not oscillate, the probability of detecting flavor $\alpha$ is
\beq
P_\alpha = (\sum_{i=1}^3 m^2_{\nu,i}|U_{\alpha i}|^2)/(\sum_{i=1}^3 m^2_{\nu,i})~.
\eeq
Using $\{m^2_{\nu,1},m^2_{\nu,2},m^2_{\nu,3}\} = \{0,7.40 \times 10^{-5},2.494 \times 10^{-3}\}$ eV$^2$, which is consistent with solar and atmospheric neutrino oscillation data \cite{1811.05487,nufit} if a normal mass hierarchy is assumed, we get
\beq
\begin{bmatrix}
P_e \\
P_\mu \\
P_\tau
\end{bmatrix}_{\rm normal} = 
\begin{bmatrix}
0.03 \\
0.52 \\
0.45
\end{bmatrix}~.
\eeq
For an inverted mass hierarchy, $\{m^2_{\nu,1},m^2_{\nu,2},m^2_{\nu,3}\} = \{2.42\times 10^{-3},2.494 \times 10^{-3},0\}$ eV$^2$ is consistent with data, and we get
\beq
\nn
\begin{bmatrix}
P_e \\
P_\mu \\
P_\tau
\end{bmatrix}_{\rm inverted} = 
\begin{bmatrix}
0.48 \\
0.24 \\
0.28
\end{bmatrix}~.
\eeq
It is also possible that the neutrino mass spectrum is heavy and near-degenerate, $m^2_{\nu,1} \simeq m^2_{\nu,2} \simeq m^2_{\nu,3}$, in which case we have
\beq
\nn
\begin{bmatrix}
P_e \\
P_\mu \\
P_\tau
\end{bmatrix}_{\rm near-degenerate} = 
\begin{bmatrix}
0.3\bar{3} \\
0.3\bar{3} \\
0.3\bar{3}
\end{bmatrix}~.
\eeq

\section{Proton recoils at BOREXINO}
\label{app:borex}

To derive our bound from \acro{borexino}, we follow the method originally developed by \cite{Beacom:2002hs,1103.2768} for detecting muon and tau neutrinos from a supernova burst.
The apparent electron equivalent recoil energy $E_R^e$ in a liquid scintillator is related to the proton recoil energy $E_R^p$ by Birks' formula for quenching:
\beq
E_{R}^e (E_R^p) = \int_0^{E_R^p} dE'^p_R \left(1 + k_{\rm B} \langle dT/dx\rangle \right)^{-1}~.
\label{eq:quenchmythirstforscintillatingliquids}
\eeq

We only consider elastic scattering on hydrogen nuclei in the organic scintillator, as scattering on carbon is greatly quenched and unobservable \cite{Beacom:2002hs,1103.2768}.
Reference \cite{1810.10543} estimated that the scattering rate per proton is restricted to be 
\bea
\nn R_p < 2 \times 10^{-39} \ {\rm s}^{-1} \ {\rm for} && \ E_{R}^e > 12.5 \ {\rm MeV} \\
\implies&& E_R^p > 19.7 \ {\rm MeV}. 
\label{eq:borexbound}
\eea

The last line of this equation is obtained by numerically solving Eq.~\eqref{eq:quenchmythirstforscintillatingliquids}
with a Birks' factor $k_{\rm B}$ = 0.011 cm/MeV \cite{1308.0443},
and by obtaining $\langle dT/dx\rangle$ (as a function of $E'^p_R$) with tables from \cite{pstar} for toluene, 
a good approximation for the \acro{borexino} scintillator, pseudocumene.

The bound on $(G^2_T/G^2_F)\sigmaveeann$ is then obtained from Eqs.~\eqref{eq:flux}, \eqref{eq:xsscat} and \eqref{eq:borexbound}, where we now use the
dipole form factor $F(q^2) = 1/(1+q^2/(0.71 \ {\rm GeV}^2))^{2}$ \cite{1711.04531} in Eq.~\eqref{eq:xsscat}.
We note that the rate of inelastic processes that may induce fragmentation or excitation of the proton is highly suppressed for the momentum transfers considered here~\cite{1711.04531}.

\input{refs.tex}
\end{document}

%% file: universalnewcommands.tex
\newcommand{\lsim}{\lesssim}
\newcommand{\ra}{\rightarrow}

\def\sigmaveeann{\langle \sigma_{\rm ann} v\rangle}

\def\Oc{\mathcal{O}}

\newcommand{\acro}[1]{\textsc{\MakeLowercase{#1}}} 
\newcommand{\osn}{\oldstylenums}
 
\newcommand{\beq}{\begin{equation}}
\newcommand{\eeq}{\end{equation}}
\newcommand{\bea}{\begin{eqnarray}}
\newcommand{\eea}{\end{eqnarray}}
\newcommand{\nn}{\nonumber}

%% file: nuCrayon_arXiv.bbl
\begin{thebibliography}{99}


\bibitem{Beacom:2006tt} 
  J.~F.~Beacom, N.~F.~Bell and G.~D.~Mack,
  Phys.\ Rev.\ Lett.\  {\bf 99}, 231301 (2007)
  [astro-ph/0608090].


\bibitem{0707.0196} 
  H.~Yuksel, S.~Horiuchi, J.~F.~Beacom and S.~Ando,
  Phys.\ Rev.\ D {\bf 76}, 123506 (2007)
  [arXiv:0707.0196 [astro-ph]].


\bibitem{1512.07501} 
  E.~Aprile {\it et al.} [XENON Collaboration],
  JCAP {\bf 1604}, no. 04, 027 (2016)
  [arXiv:1512.07501 [physics.ins-det]].
  
  \bibitem{1509.02910} 
  D.~S.~Akerib {\it et al.} [LZ Collaboration],
  arXiv:1509.02910 [physics.ins-det].


\bibitem{1606.07001} 
  J.~Aalbers {\it et al.} [DARWIN Collaboration],
  JCAP {\bf 1611}, 017 (2016)
  [arXiv:1606.07001 [astro-ph.IM]].


\bibitem{1707.08145} 
  C.~E.~Aalseth {\it et al.},
  Eur.\ Phys.\ J.\ Plus {\bf 133}, 131 (2018)
  [arXiv:1707.08145 [physics.ins-det]].


\bibitem{largo}
		\url{https://indico.in2p3.fr/event/17777/}, \url{http://deap3600.ca/deap-50t-detector/}, \url{https://goo.gl/re849A}
		
\bibitem{1803.08044} 
  J.~Bramante, B.~Broerman, R.~F.~Lang and N.~Raj,
  Phys.\ Rev.\ D {\bf 98}, no. 8, 083516 (2018)
  [arXiv:1803.08044 [hep-ph]].
		
\bibitem{1711.05283} 
  A.~Olivares-Del Campo, C.~Boehm, S.~Palomares-Ruiz and S.~Pascoli,
  Phys.\ Rev.\ D {\bf 97}, no. 7, 075039 (2018)
  [arXiv:1711.05283 [hep-ph]].


\bibitem{1604.01218} 
  K.~Abe {\it et al.} [XMASS Collaboration],
  Astropart.\ Phys.\  {\bf 89}, 51 (2017)
  [arXiv:1604.01218 [physics.ins-det]].


\bibitem{1606.09243} 
  R.~F.~Lang, C.~McCabe, S.~Reichard, M.~Selvi and I.~Tamborra,
  Phys.\ Rev.\ D {\bf 94}, no. 10, 103009 (2016)
  [arXiv:1606.09243 [astro-ph.HE]].


\bibitem{1806.05651} 
  M.~B.~Voloshin,
  Phys.\ Rev.\ D {\bf 98}, no. 3, 034025 (2018)
  [arXiv:1806.05651 [hep-ph]].


\bibitem{1806.01417} 
  T.~Kozynets, S.~Fallows and C.~B.~Krauss,
  Astropart.\ Phys.\  {\bf 105}, 25 (2019)
  [arXiv:1806.01417 [astro-ph.HE]].


\bibitem{1506.08309} 
  M.~Schumann, L.~Baudis, L.~Bütikofer, A.~Kish and M.~Selvi,
  JCAP {\bf 1510}, no. 10, 016 (2015)
  [arXiv:1506.08309 [physics.ins-det]].


\bibitem{1510.04196} 
  D.~Franco {\it et al.},
  JCAP {\bf 1608}, no. 08, 017 (2016)
  [arXiv:1510.04196 [physics.ins-det]].
  
    
  \bibitem{Cerdeno:2016sfi} 
  D.~G.~Cerdeño, M.~Fairbairn, T.~Jubb, P.~A.~N.~Machado, A.~C.~Vincent and C.~B½hm,
  JHEP {\bf 1605}, 118 (2016)
  Erratum: [JHEP {\bf 1609}, 048 (2016)]
  [arXiv:1604.01025 [hep-ph]].

\bibitem{1807.07169} 
  J.~L.~Newstead, L.~E.~Strigari and R.~F.~Lang,
  arXiv:1807.07169 [astro-ph.SR].


\bibitem{1711.04531} 
  Y.~Cui, M.~Pospelov and J.~Pradler,
  Phys.\ Rev.\ D {\bf 97}, no. 10, 103004 (2018)
  [arXiv:1711.04531 [hep-ph]].


\bibitem{1501.03166} 
  J.~F.~Cherry, M.~T.~Frandsen and I.~M.~Shoemaker,
  Phys.\ Rev.\ Lett.\  {\bf 114}, 231303 (2015)
  [arXiv:1501.03166 [hep-ph]].


\bibitem{Bertoni:2014mva} 
  B.~Bertoni, S.~Ipek, D.~McKeen and A.~E.~Nelson,
  JHEP {\bf 1504}, 170 (2015)
  [arXiv:1412.3113 [hep-ph]].


\bibitem{Batell:2017rol} 
  B.~Batell, T.~Han and B.~Shams Es Haghi,
  Phys.\ Rev.\ D {\bf 97}, no. 9, 095020 (2018)
  [arXiv:1704.08708 [hep-ph]].


\bibitem{Batell:2017cmf} 
  B.~Batell, T.~Han, D.~McKeen and B.~Shams Es Haghi,
  Phys.\ Rev.\ D {\bf 97}, no. 7, 075016 (2018)
  [arXiv:1709.07001 [hep-ph]].


\bibitem{1112.4491} 
  G.~Belanger and J.~C.~Park,
  JCAP {\bf 1203}, 038 (2012)
  [arXiv:1112.4491 [hep-ph]].


\bibitem{1312.0011} 
  J.~Huang and Y.~Zhao,
  JHEP {\bf 1402}, 077 (2014)
  [arXiv:1312.0011 [hep-ph]].


\bibitem{1405.7370} 
  K.~Agashe, Y.~Cui, L.~Necib and J.~Thaler,
  JCAP {\bf 1410}, no. 10, 062 (2014)
  [arXiv:1405.7370 [hep-ph]].


\bibitem{1410.2246} 
  J.~Berger, Y.~Cui and Y.~Zhao,
  JCAP {\bf 1502}, no. 02, 005 (2015)
  [arXiv:1410.2246 [hep-ph]].


\bibitem{1411.6632} 
  K.~Kong, G.~Mohlabeng and J.~C.~Park,
  Phys.\ Lett.\ B {\bf 743}, 256 (2015)
  [arXiv:1411.6632 [hep-ph]].


\bibitem{1503.02669} 
  J.~Kopp, J.~Liu and X.~P.~Wang,
  JHEP {\bf 1504}, 105 (2015)
  [arXiv:1503.02669 [hep-ph]].


\bibitem{1610.03486} 
  L.~Necib, J.~Moon, T.~Wongjirad and J.~M.~Conrad,
  Phys.\ Rev.\ D {\bf 95}, no. 7, 075018 (2017)
  [arXiv:1610.03486 [hep-ph]].


\bibitem{1611.09866} 
  H.~Alhazmi, K.~Kong, G.~Mohlabeng and J.~C.~Park,
  JHEP {\bf 1704}, 158 (2017)
  [arXiv:1611.09866 [hep-ph]].


\bibitem{1612.02834} 
  A.~Bhattacharya, R.~Gandhi, A.~Gupta and S.~Mukhopadhyay,
  JCAP {\bf 1705}, no. 05, 002 (2017)
  [arXiv:1612.02834 [hep-ph]].


\bibitem{1612.06867} 
  D.~Kim, J.~C.~Park and S.~Shin,
  Phys.\ Rev.\ Lett.\  {\bf 119}, no. 16, 161801 (2017)
  [arXiv:1612.06867 [hep-ph]].


\bibitem{1711.05278} 
  C.~Kachulis {\it et al.} [Super-Kamiokande Collaboration],
  Phys.\ Rev.\ Lett.\  {\bf 120}, no. 22, 221301 (2018)
  [arXiv:1711.05278 [hep-ex]].


\bibitem{1712.07126} 
  G.~F.~Giudice, D.~Kim, J.~C.~Park and S.~Shin,
  Phys.\ Lett.\ B {\bf 780}, 543 (2018)
  [arXiv:1712.07126 [hep-ph]].


\bibitem{1803.03264} 
  A.~Chatterjee, A.~De Roeck, D.~Kim, Z.~G.~Moghaddam, J.~C.~Park, S.~Shin, L.~H.~Whitehead and J.~Yu,
  Phys.\ Rev.\ D {\bf 98}, no. 7, 075027 (2018)
  [arXiv:1803.03264 [hep-ph]].


\bibitem{1804.07302} 
  D.~Kim, K.~Kong, J.~C.~Park and S.~Shin,
  JHEP {\bf 1808}, 155 (2018)
  [arXiv:1804.07302 [hep-ph]].


\bibitem{1806.09154} 
  M.~Aoki and T.~Toma,
  JCAP {\bf 1810}, no. 10, 020 (2018)
  [arXiv:1806.09154 [hep-ph]].


\bibitem{1811.09344} 
  C.~Ha {\it et al.},
  arXiv:1811.09344 [astro-ph.IM].
  
  \bibitem{0907.0018} 
  R.~Catena and P.~Ullio,
  JCAP {\bf 1008}, 004 (2010)
  [arXiv:0907.0018 [astro-ph.CO]].


\bibitem{1012.4515} 
  M.~Cirelli {\it et al.},
  JCAP {\bf 1103}, 051 (2011)
  Erratum: [JCAP {\bf 1210}, E01 (2012)]
  [arXiv:1012.4515 [hep-ph]].


\bibitem{Navarro:1995iw} 
  J.~F.~Navarro, C.~S.~Frenk and S.~D.~M.~White,
  Astrophys.\ J.\  {\bf 462}, 563 (1996)
  [astro-ph/9508025].


\bibitem{1204.3622} 
  G.~Steigman, B.~Dasgupta and J.~F.~Beacom,
  Phys.\ Rev.\ D {\bf 86}, 023506 (2012)
  [arXiv:1204.3622 [hep-ph]].


\bibitem{1602.07282} 
  A.~Moline, J.~A.~Schewtschenko, S.~Palomares-Ruiz, C.~B½hm and C.~M.~Baugh,
  JCAP {\bf 1608}, 069 (2016)
  [arXiv:1602.07282 [astro-ph.CO]].


\bibitem{1809.00671} 
  N.~Klop and S.~Ando,
  Phys.\ Rev.\ D {\bf 98}, no. 10, 103004 (2018)
  [arXiv:1809.00671 [hep-ph]].
    
  
\bibitem{Bertone:2005xz} 
  G.~Bertone, A.~R.~Zentner and J.~Silk,
  Phys.\ Rev.\ D {\bf 72}, 103517 (2005)
  [astro-ph/0509565].


\bibitem{Bertone:2006nq} 
  G.~Bertone,
  Phys.\ Rev.\ D {\bf 73}, 103519 (2006)
  [astro-ph/0603148].


\bibitem{Horiuchi:2006de} 
  S.~Horiuchi and S.~Ando,
  Phys.\ Rev.\ D {\bf 74}, 103504 (2006)
  [astro-ph/0607042].


\bibitem{Brun:2007tn} 
  P.~Brun, G.~Bertone, J.~Lavalle, P.~Salati and R.~Taillet,
  Phys.\ Rev.\ D {\bf 76}, 083506 (2007)
  [arXiv:0704.2543 [astro-ph]].
  
  
  \bibitem{Arina:2015zoa} 
  C.~Arina, S.~Kulkarni and J.~Silk,
  Phys.\ Rev.\ D {\bf 92}, no. 8, 083519 (2015)
  [arXiv:1506.08202 [astro-ph.HE]].


\bibitem{Freedman:1977xn} 
  D.~Z.~Freedman, D.~N.~Schramm and D.~L.~Tubbs,
  Ann.\ Rev.\ Nucl.\ Part.\ Sci.\  {\bf 27}, 167 (1977).


\bibitem{Erler:2004in} 
  J.~Erler and M.~J.~Ramsey-Musolf,
  Phys.\ Rev.\ D {\bf 72}, 073003 (2005)
  [hep-ph/0409169].


\bibitem{1103.3261} 
  M.~Pospelov,
  Phys.\ Rev.\ D {\bf 84}, 085008 (2011)
  [arXiv:1103.3261 [hep-ph]].

\bibitem{Helm:1956zz} 
  R.~H.~Helm,
  Phys.\ Rev.\  {\bf 104}, 1466 (1956).
  
  \bibitem{private}
  P.~Giampa, private communication.

\bibitem{0903.3630} 
  L.~E.~Strigari,
  New J.\ Phys.\  {\bf 11}, 105011 (2009)
  [arXiv:0903.3630 [astro-ph.CO]].


\bibitem{1307.5458} 
  J.~Billard, L.~Strigari and E.~Figueroa-Feliciano,
  Phys.\ Rev.\ D {\bf 89}, no. 2, 023524 (2014)
  [arXiv:1307.5458 [hep-ph]].


\bibitem{1805.12562} 
  E.~Aprile {\it et al.} [XENON Collaboration],
  Phys.\ Rev.\ Lett.\  {\bf 121}, no. 11, 111302 (2018)
  [arXiv:1805.12562 [astro-ph.CO]].


\bibitem{0911.0548} 
  G.~Bellini {\it et al.} [Borexino Collaboration],
  Phys.\ Rev.\ C {\bf 81}, 034317 (2010)
  [arXiv:0911.0548 [hep-ex]].


\bibitem{1105.3516} 
  A.~Gando {\it et al.} [KamLAND Collaboration],
  Astrophys.\ J.\  {\bf 745}, 193 (2012)
  [arXiv:1105.3516 [astro-ph.HE]].


\bibitem{1801.10159} 
  R.~Essig, M.~Sholapurkar and T.~T.~Yu,
  Phys.\ Rev.\ D {\bf 97}, no. 9, 095029 (2018)
  [arXiv:1801.10159 [hep-ph]].


\bibitem{1111.5031} 
  K.~Bays {\it et al.} [Super-Kamiokande Collaboration],
  Phys.\ Rev.\ D {\bf 85}, 052007 (2012)
  [arXiv:1111.5031 [hep-ex]].
  
  
  \bibitem{1701.07209} 
  C.~Garcia-Cely and J.~Heeck,
  JHEP {\bf 1705}, 102 (2017)
  [arXiv:1701.07209 [hep-ph]].
    
  \bibitem{0711.4866} 
  M.~Pospelov, A.~Ritz and M.~B.~Voloshin,
  Phys.\ Lett.\ B {\bf 662}, 53 (2008)
  [arXiv:0711.4866 [hep-ph]].
  
  \bibitem{Aarssen:2012fx} 
  L.~G.~van den Aarssen, T.~Bringmann and C.~Pfrommer,
  Phys.\ Rev.\ Lett.\  {\bf 109}, 231301 (2012)
  [arXiv:1205.5809 [astro-ph.CO]].
  
  \bibitem{Shoemaker:2013tda} 
  I.~M.~Shoemaker,
  Phys.\ Dark Univ.\  {\bf 2}, no. 3, 157 (2013)
  [arXiv:1305.1936 [hep-ph]].
  
  \bibitem{Ko:2014bka} 
  P.~Ko and Y.~Tang,
  Phys.\ Lett.\ B {\bf 739}, 62 (2014)
  [arXiv:1404.0236 [hep-ph]].
  
  \bibitem{Archidiacono:2014nda} 
  M.~Archidiacono, S.~Hannestad, R.~S.~Hansen and T.~Tram,
  Phys.\ Rev.\ D {\bf 91}, no. 6, 065021 (2015)
  [arXiv:1404.5915 [astro-ph.CO]].
  
  \bibitem{Cherry:2014xra} 
  J.~F.~Cherry, A.~Friedland and I.~M.~Shoemaker,
  arXiv:1411.1071 [hep-ph].

      
  \bibitem{Falkowski:2009yz} 
  A.~Falkowski, J.~Juknevich and J.~Shelton,
  arXiv:0908.1790 [hep-ph].
  
  
\bibitem{1311.5764} 
  M.~Pospelov and J.~Pradler,
  Phys.\ Rev.\ D {\bf 89}, no. 5, 055012 (2014)
  [arXiv:1311.5764 [hep-ph]].
  
  \bibitem{Kaplinghat:2000vt} 
  M.~Kaplinghat, L.~Knox and M.~S.~Turner,
  Phys.\ Rev.\ Lett.\  {\bf 85}, 3335 (2000)
  doi:10.1103/PhysRevLett.85.3335
  [astro-ph/0005210].


 
  
  \bibitem{Barger:2007hj} 
  V.~D.~Barger, W.~Y.~Keung and G.~Shaughnessy,
  Phys.\ Lett.\ B {\bf 664}, 190 (2008)
  doi:10.1016/j.physletb.2008.05.021
  [arXiv:0709.3301 [astro-ph]].
  

   \bibitem{1609.04876} 
  C.~Rott, S.~In, J.~Kumar and D.~Yaylali,
  JCAP {\bf 1701}, no. 01, 016 (2017)
  [arXiv:1609.04876 [hep-ph]].
  
\bibitem{Bell:2014xta} 
  N.~F.~Bell, S.~Horiuchi and I.~M.~Shoemaker,
  Phys.\ Rev.\ D {\bf 91}, no. 2, 023505 (2015)
  [arXiv:1408.5142 [hep-ph]].
  
  \bibitem{Goodman:1984dc} 
  M.~W.~Goodman and E.~Witten,
  Phys.\ Rev.\ D {\bf 31}, 3059 (1985).
  
  \bibitem{Drukier:1983gj} 
  A.~Drukier and L.~Stodolsky,
  Phys.\ Rev.\ D {\bf 30}, 2295 (1984).

  
  \bibitem{1811.05487} 
  I.~Esteban, M.~C.~Gonzalez-Garcia, A.~Hernandez-Cabezudo, M.~Maltoni and T.~Schwetz,
  arXiv:1811.05487 [hep-ph].
  
  \bibitem{nufit}
  \url{http://www.nu-fit.org/}


\bibitem{Beacom:2002hs} 
  J.~F.~Beacom, W.~M.~Farr and P.~Vogel,
  Phys.\ Rev.\ D {\bf 66}, 033001 (2002)
  [hep-ph/0205220].


\bibitem{1103.2768} 
  B.~Dasgupta and J.~F.~Beacom,
  Phys.\ Rev.\ D {\bf 83}, 113006 (2011)
  [arXiv:1103.2768 [hep-ph]].


\bibitem{1810.10543} 
  T.~Bringmann and M.~Pospelov,
  arXiv:1810.10543 [hep-ph].


\bibitem{1308.0443} 
  G.~Bellini {\it et al.} [Borexino Collaboration],
  Phys.\ Rev.\ D {\bf 89}, no. 11, 112007 (2014)
  [arXiv:1308.0443 [hep-ex]].


\bibitem{pstar}
		\url{https://physics.nist.gov/PhysRefData/Star/Text/PSTAR.html}
		
		
		
\end{thebibliography}
